\documentclass[fleqn,usenatbib]{mnras}

\usepackage{newtxtext,newtxmath}

\usepackage[T1]{fontenc}

\DeclareRobustCommand{\VAN}[3]{#2}
\let\VANthebibliography\thebibliography
\def\thebibliography{\DeclareRobustCommand{\VAN}[3]{##3}\VANthebibliography}

\usepackage{graphicx}	
\usepackage{amsmath}	
\usepackage{makecell}

\defcitealias{Galloway2008}{1}
\defcitealias{intZand2007}{2}
\defcitealias{Linares20124U}{3}
\defcitealias{Jenke2016}{4}
\defcitealias{papitto2024}{5}
\defcitealias{Linares2010}{6}
\defcitealias{Bagnoli2013}{7}
\defcitealias{Linares2012}{8}
\defcitealias{Falanga2011}{9}
\defcitealias{wang2024}{10}
\defcitealias{Galloway2004}{11}
\defcitealias{li2018}{12}
\defcitealias{Fiocchi2019}{46}

\defcitealias{tse2023}{12}
\defcitealias{revnivtsev2001}{13}
\defcitealias{strohmayer2011}{14}
\defcitealias{mancuso2019}{15}
\defcitealias{mancuso2023}{16}
\defcitealias{strohmayer2018}{17}
\defcitealias{tse2021}{18}
\defcitealias{buisson2020}{19}
\defcitealias{molkov2005}{43}

\defcitealias{Schaefer_2010}{20}
\defcitealias{Selvelli2008}{21}
\defcitealias{patterson2017}{22}
\defcitealias{Patterson2022}{23}
\defcitealias{lederle2003}{24}
\defcitealias{Sion2019}{25}
\defcitealias{rodriguez2023}{26}
\defcitealias{schaefer2022V2487}{27}
\defcitealias{Schaefer1995}{28}
\defcitealias{thoroughgood2001}{29}
\defcitealias{stanishev2004}{30}
\defcitealias{Zamanov2023}{31}
\defcitealias{brandi2009}{32}
\defcitealias{mikolajewska2017}{33}
\defcitealias{nelson2011}{34}
\defcitealias{Schaefer2009}{35}
\defcitealias{mikolajewska2021}{36}
\defcitealias{Schaefer2022}{37}
\defcitealias{darnley2016}{38}
\defcitealias{darnley2017}{39}
\defcitealias{shore1991}{40}
\defcitealias{kuin2020}{41}
\defcitealias{bode2016}{42}
\defcitealias{shafter2022m31n}{43}
\defcitealias{shafter2022}{44}
\defcitealias{shafter2024}{45}

\title[$\dot m$ and $t_{\rm rec}$ on NSs and WDs]{Connecting the m-dots: accretion rates and thermonuclear burst recurrence times on neutron stars and white dwarfs}

\author[Triantafyllos Kormpakis, Manuel Linares, Jordi Jos\'e]{
Triantafyllos Kormpakis, $^{1,}$$^{2}$\thanks{E-mail: triantafyllos.kormpakis@ntnu.no}
Manuel Linares,$^{1,}$$^{2}$
Jordi Jos\'e$^{2,}$$^{3}$
\\
$^{1}$Department of Physics, Norwegian University of Science and Technology, NO-7491 Trondheim, Norway\\
$^{2}$Departament de Física, EEBE, Universitat Polit\`ecnica de Catalunya, Av. Eduard Maristany 16, E-08019 Barcelona, Spain\\
$^{3}$Institut d’Estudis Espacials de Catalunya, c/Esteve Terradas 1, E-08860 Castelldefels, Spain
}

\date{Accepted XXX. Received YYY; in original form ZZZ}

\pubyear{2024}

\begin{document}
\label{firstpage}
\pagerange{\pageref{firstpage}--\pageref{lastpage}}
\maketitle


\begin{abstract}
    We present a compilation of observed recurrence times ($t_{\rm rec}$) and infer the corresponding local mass-accretion rates ($\dot m$) for type I X-ray bursts, milliHertz quasi-periodic oscillating sources and recurrent novae eruptions. We construct models of the $t_{\rm rec}-\dot m$ relation for accreting white dwarfs and neutron stars and find that both are roughly consistent with a global inverse linear relation, connecting for the first time thermonuclear runaways on neutron stars and white dwarfs. We find that theoretical models of pure He bursts are in agreement with the best $t_{\rm rec}$ measurements in ultra-compact X-ray binaries at low $\dot m$ (4U~$0614+09$ and 2S~0918-549).
    We suggest that the transient Z source XTE~J1701-462 is a slow rotator, based on its mHz QPO properties. Finally, we discuss the implications for thermonuclear ignition and point out that the difference in eruption/burst energy ($E_{b_{WD}}/E_{b_{NS}}=2\times 10^4$) is consistent with the difference in area between neutron stars and white dwarfs $\left((R_{WD}/R_{NS})^2=4\times 10^4\right)$. We conclude that ignitions of thermonuclear shell flashes on neutron stars and white dwarfs depend primarily on the specific mass accretion rate and do not depend on the nature of the underlying compact object.
\end{abstract}

\begin{keywords}
accretion -- X-rays: bursts -- novae, cataclysmic variables.
\end{keywords}



\section{Introduction}

Neutron stars (NSs) and white dwarfs (WDs) in mass-transferring binaries accrete hydrogen- or helium-rich matter from the surfaces of their stellar companions. Depending on the 
mass-accretion rate $\dot{M}$, nuclear burning in the accreted fuel can be thermally stable or unstable (\citealt{Fujimoto81,nomoto1982accreting, narayan2003thermonuclear}), or even marginally stable in some cases \citep{heger2007}. Unstable burning of the accreted material in compact binaries gives rise to thermonuclear runaways (TNRs), where the accumulated material ignites at a specific depth in the envelope of the NS or WD, releasing energy and being observed as a shell flash. In NSs, these TNRs are seen mostly in X-rays and are known as {\it type I X-ray bursts} \citep[XRBs hereafter,][]{lewin1993}.
TNRs on WDs release most of their energy in the optical band and are commonly named {\it nova eruptions} (in the systems known as classical or recurrent novae, CNe or RNe, respectively; \citealt{munari2012classical,pagnotta2014}).
XRBs give us the ability to constrain the NS mass and radius, mainly through photospheric radius expansion (PRE) bursts \citep{van1979,paczynski1986}.
In nova eruptions, determining the mass and composition of the ejecta provides us with information on the chemical enrichment of the interstellar medium. More importantly, RNe are possible progenitors of type Ia supernovae \citep{hachisu1996}, since the WD mass is expected to increase in each nova cycle.

The recurrence times ($t_\mathrm{rec}$\footnote{We use the same symbol $t_{\rm rec}$ for NS bursts and WD eruptions.})
of XRBs are connected to the accretion rate onto the NS, and range from minutes \citep{Linares2012} to years \citep{Kuulkers04}.
Because a thicker fuel layer takes longer to accumulate and cools down more slowly after burning, more energetic bursts have longer $t_{\rm rec}$ and durations \citep[see, e.g.,][for reviews]{Strohmayer06,jose2016stellar}. 
Normal H/He bursts on NSs (lasting from seconds to minutes) release an energy $E_b \sim 10^{38}-10^{40}$~erg and have typical $t_\mathrm{rec}$ ranging from minutes to days \citep{galloway2020}.
Unusually long or "intermediate" bursts (lasting tens of minutes) release $E_{\rm b} \sim 10^{40}-10^{41}$~erg and have $t_\mathrm{rec}$ of weeks to months \citep{cumming2006,Alizai23}.
While overlaps between long and normal bursts exist \citep{Linares20124U}, long bursts are thought to be produced by the ignition of a thick layer of pure He \citep{cumming2006,intZand2007,Peng07}.
Superbursts (lasting from hours to about a day) release $E_{\rm b} \sim 10^{42}$~erg and have $t_\mathrm{rec}$ of years \citep{Cornelisse00,Kuulkers04,Zand2017}.
Because they are attributed to the ignition of a different element \citep[carbon;][]{Cumming01b,Strohmayer2002}, superbursts are not included in this study.

Nova eruptions are produced by H shell flashes on WDs \citep{starrfield1972cno}, with observed $t_{\rm rec}$ between one year and several decades for RNe \citep{Schaefer_2010}.
For CNe, by definition, only one eruption has been observed so it is not possible to measure directly $t_{\rm rec}$.
However, CNe models typically predict\footnote{\cite{Yaron2005} reported $P_{\rm rec}$ for a wide range of models, obtaining $P_{\rm rec}$ values between $0.0178$ yr and  $1.08 \times 10^9$ yr. The most extreme values of $P_{\rm rec}$ are only reachable for extreme combinations of $M_{\rm WD}$ and $M_{\rm acc}$.} $t_{\rm rec} \sim 10^4-10^5$~yr \citep[e.g.,][]{Shara_2018}.
The ejected envelope expands into the surrounding material powering most of the optical transient emission, which typically lasts for days to months (\citealt{chomiuk2021new}; see, e.g., decline times of 93 CNe in the range $1 - 295$~d, \citealt{strope2010}). In RNe, the flash duration (initial brightening to final decline) is observationally found to be $19 - 429$~d \citep{Schaefer_2010}.
The energy released in a nova explosion is typically $10^{44}-10^{45}$~erg (see \citealt{starrfield2008thermonuclear,jose2008,jose2016stellar,chomiuk2021new}, for reviews).
%

The main parameters considered in ignition models for XRBs are: i) the local mass-accretion rate on the NS surface (per unit surface area, $\dot{m}$); ii) the accreted fuel composition (mass fractions of H, He and CNO elements: $X_0$, $Y_0$ and $Z$, respectively);
and iii) the energy released from the deep crust, presumably from pycnonuclear reactions, heating up the accreted envelope "from below" \citep{Fujimoto81,bildsten1998,Bildsten2000,haensel2003}. 
When accreting solar composition fuel at the lowest $\dot{m}$ ($\lesssim$1\% of the Eddington rate\footnote{$\dot M_{\rm Edd}=\frac{L_{\rm Edd}}{\eta c^2}$, where $L_{\rm Edd}$ is the Eddington limited luminosity, $\eta$ is the radiative efficiency, and c the speed of light.}), unstable H burning can trigger a combined H/He burst \citep[``H bursts'',][]{Peng07}.
At higher rates (about 1$-$4\% Eddington), all H burns stably before reaching ignition depth, so that a thick He layer ignites in the absence of H \citep[``pure He bursts",][]{Fujimoto81,Galloway06b}.
NSs in ultra-compact binaries (UCXBs, with orbital periods shorter than 1~h) are thought to accrete mainly He, which can also lead to (more frequent and less energetic) ``pure He bursts'' \citep{Cumming03,Peng10}.
Most XRBs at relatively high accretion rates (about 4$-$100\% Eddington) are thought to be triggered by He ignition in a mixed H and He environment, since steady H burning cannot exhaust the accreted H before reaching ignition depth \citep[``mixed H/He bursts'',][]{bildsten1998}.
Finally, for very high accretion rates onto NSs
(close to or above the Eddington limit)
both H and He are expected to burn steadily \citep{bildsten1998,Bildsten2000}.

When the accretion rate onto the NS is close to the boundary between unstable and stable He burning, an oscillatory phenomenon called marginally-stable burning occurs \citep[with a period similar to the geometric mean of thermal and accretion timescales;][]{heger2007}.
This is the leading model for the {\it mHz quasi-periodic oscillations} (mHz QPOs) observed in the X-ray flux of several accreting NSs
\citep[][see Section \ref{sec:mhz}]{revnivtsev2001}.
Thus, mHz QPOs can be used as a tracer for the stability boundary of He burning on accreting NSs. 

The main parameters of the models for H ignition on WDs are, similarly to NSs: i) the $\dot M$, ii) the $M_{\rm {WD}}$, iii) the WD temperature (or luminosity) and iv) metallicity $Z$ of the accreted matter \citep{wolf2013,Yaron2005,Shara_2018}.
\citet{Bildsten2007} proposed that He novae can occur in a particular sub-class of WD binaries known as AM Canum Venaticorum (AM CVn), where unstable He burning on the accreting WD would give rise to He flashes (see also \citealt{shen2009}).
In the case of V445 Puppis, while the spectra lack H and are rich in CI lines, do not conclusively point to a Helium nova  \citep{Woudt_2009,ashok2003}. So far, only a handful of potential He novae have been proposed \citep{Rosenbush2008}.


Steady burning of H-rich matter is thought to power the super-soft phase of nova eruptions, and the so-called super-soft sources \citep[SSSs,][]{Kahabka97}.
These are thought to accrete at a rate $\dot M = 10^{-7}$ $\rm{M_{\odot}}$ $\rm{yr^{-1}}$, where H burning on a WD is expected to be stable \citep{shen2007}.
Recently, a peculiar SSS was identified (lacking H lines), which could be burning He stably \citep{greiner2023}.
Thus, SSSs can help us trace the stability boundary of H (and perhaps He) burning on accreting WDs.

Here we investigate the relation between $t_\mathrm{rec}$ and $\dot{M}$, comparing observations of TNRs on NSs and WDs with models, in order to study H and He burning regimes on different compact objects.
%
In Section \ref{sec:NS} we compile XRB and mHz QPO properties and in Section \ref{sec:WDtrec} we collect $t_\mathrm{rec}$ and $\dot{M}$ for nova eruptions in RNe.
In Section~\ref{sec:Models} we describe the ignition models used in this work, 
and in Section \ref{sec:Discussion} we discuss our results.

\section{Neutron Stars}
\label{sec:NS}

To compare thermonuclear burning regimes in NSs and WDs, we gathered from the literature estimates of $t_\mathrm{rec}$ and the corresponding $\dot{m}$. 
We define $t_\mathrm{rec}$ as the time elapsed between successive bursts in sequences of many bursts (more than 3).
Exceptions of burst trains with short ``wait times'' exist, but are not included in this definition \citep{Keek2010}.
In practice, for $t_\mathrm{rec} \gtrsim 1$~d it is not possible to observe an XRB source (or ``burster'') uninterruptedly, so we must measure the long-term average $t_\mathrm{rec}$ \citep[number of detected bursts over the effective exposure or on-source time; see, e.g.,][]{Linares20124U,Jenke2016}.
In a few cases, $t_\mathrm{rec}$ can be studied over a range of $\dot{m}$ for one particular burster; most notably: GS~1826$-$24, IGR~J17511$-$3057 and T5X2 \citep[][respectively]{Galloway2004,Falanga2011,Linares2012}.
In most bursters we must resort to large multi-instrument X-ray observations to constrain $t_\mathrm{rec}$ averaged over long timescales \citep[years-decade;][]{intZand2007,Galloway2008,galloway2020}.
In addition, the accretion rate and luminosity can change strongly (by several orders of magnitude) in low-mass X-ray binaries (LMXBs), due to accretion disk instabilities (both in persistent and, especially, transient accretors).
Thus, for several bursters we list and plot a range of $\dot{m}$ (Sec.~\ref{sec:NStrec}).

XRBs come from weakly magnetic (B$\lesssim$10$^{10}$~G) accreting NSs in LMXBs, most of them in transient systems. 
Their persistent X-ray luminosity (outside bursts, $L_X$) can be used to estimate the ``bolometric'' accretion luminosity $L_{\rm acc} = c_{\rm bol}\times L_X$ and $\dot M$ \citep[e.g.,][]{Galloway2008}.
Here $c_{\rm bol}$ is a bolometric correction factor that converts from X-ray luminosity $L_X$ in a given observed band (set by the different detectors used) to the band where most accretion-powered energy is radiated (about 0.1-100~keV).
However, the exact band depends on the source state and spectral model of choice so we find different values in the literature, starting at 0.01-0.8~keV and reaching up to 50-200~keV.
Using the distances to each burster tabulated below ($D$), we calculate when necessary $L_X = 4\pi D^2 F_X$ from the reported unabsorbed X-ray flux in the same band ($F_X$).
In a few cases we estimate $F_X$ converting from instrument count rates and assuming a spectral model, as detailed below. 

We then estimate $\dot{M}$ under the simple assumption that all gravitational potential energy is released at the NS surface in the form of X-rays, using the general relativistic expression \citep{lewin1993},
\begin{equation}
\label{eq:lma}
    \dot M = \frac{L_{\rm acc} (1+z)^2}{z c^2}. 
\end{equation}
Here $1 + z = 1/\sqrt{1-2GM/Rc^2}$ is the gravitational redshift factor, and we adopt for simplicity a NS mass and radius $M$=1.4~M$_\odot$ and $R$=10~km, respectively.
%
Assuming that the accreted matter is evenly spread on the NS surface, we estimate the mass-accretion rate per unit area as simply $\dot{m} = \dot{M}/(4\pi R^2)$.
We also use $L_{\rm Edd} = 2.5 \times 10^{38}$ erg $s^{-1}$, to calculate Eddington-normalized accretion luminosities and $\dot m$ \citep[in general this value depends on $M$, $R$ and $X$; see, e.g.,][]{lewin1993,kuulkers03}.
With our definition, the Eddington limit corresponds to global and local mass-accretion rates of $\dot M_{\rm Edd} = 1.6 \times 10^{18}$~g~s$^{-1} = 2.5 \times 10^{-8} M_\odot$~yr$^{-1}$ and $\dot m_{\rm Edd} = 1.2 \times 10^5$~g~cm$^{-2}$~s$^{-1}$, respectively.
Finally, we estimate a systematic uncertainty on the general bolometric correction, using \textsc{WebPIMMS}\footnote{\url{https://heasarc.gsfc.nasa.gov/cgi-bin/Tools/w3pimms/w3pimms.pl}} and X-ray photon indices $\Gamma=1.5$ and $\Gamma=3$.
Assuming that this photon index range represents approximately the different spectral states of NS-LMXBs \citep{linares2009}, we arrive at a systematic uncertainty on $L_{\rm acc}$ (excluding $D$ uncertainties) of about $30\%$.

\subsection{The XRB sample: recurrence times and accretion rates}
\label{sec:NStrec} 

We compiled a catalog of $t_\mathrm{rec}$ and inferred $\dot{m}$ for the Galactic burster population.
From the 120 bursters currently known in the Milky Way \citep{ZandBursters}, we include those that have: i) good measurements of $t_\mathrm{rec}$; ii) known or well-constrained distances and iii) reported or available measurements of the persistent flux around the bursts. This leaves us with 50 bursters, shown in Table~\ref{tab:NS} and Figure~\ref{fig:NStrec}. The corresponding $L_{\rm acc}, c_{\rm bol}$ and $D$ values for each burster are also listed, together with the type or classification of each burster.
We also give the orbital period $P_{\rm orb}$ for each system, to distinguish between UCXBs ($P_{\rm orb} \leq 1$ h) and normal LMXBs.
This distinction serves as an indicator for the composition of the accreted fuel, since in UCXBs the NSs are expected to accrete H-poor matter.

We used two main references as a starting point to compile this information. First, \cite{intZand2007} reported $t_\mathrm{rec}$ constraints from BeppoSAX Wide Field Camera (WFC) observations of 40 persistent bursters (in their Table 2). From these 40 bursters, 28 comply with our criteria i) and ii) mentioned above, so we include them in our catalog (noted with instrument "WFC" in our Table~\ref{tab:NS}). 
The $\dot{m}$ estimates from \cite{intZand2007} come from the following instruments:
{\it RXTE} Proportional Counter Array (PCA) bulge-scan light curves (in the energy range of 2-12 keV), converted to $F_X$ in the 2-10 keV band;
and {\it RXTE} All-Sky Monitor (ASM) light curves (2-12 keV), converted to $F_X$ in the 2-10 keV band.
From there, they assume a unique bolometric correction factor 2.9($\pm$1.4) to convert to ``bolometric''
energy flux.
We also inferred the distance $D$ to each burster listed by \cite{intZand2007}, using their burst peak flux and $L_{\rm Edd}$ (these inferred $D$ values are also shown in Table~\ref{tab:NS}).

Second, we include constraints on $t_{\rm rec}$ for 12 bursters from Table 10 in \cite{Galloway2008}, which lists the shortest burst interval longer than 0.9 h (thus excluding the atypical short recurrence-time burst ``trains''). 
For completeness, we list in our Table~\ref{tab:NS} their distances $D$ (from PRE bursts, cf. their Table 9) and their $c_{\rm bol}$ values, which they derived from spectral fits to the persistent emission  to convert from the PCA band (2.5-25 keV) to ``bolometric'' 0.1-200 keV energy flux. 
Together with the two references above, we included constraints on $t_{\rm rec}$ from \textsc{MINBAR} \citep{galloway2020}, which we show in Figure~\ref{fig:NStrec}.
Finally, we also included improved or more recent $t_\mathrm{rec}$ measurements from 10
additional bursters which fulfill our selection criteria, briefly summarized next.

\subsection*{T5X2}

The burster, transient and 11~Hz X-ray pulsar in the globular cluster Terzan 5, IGR J17480-2446, was nicknamed T5X2 for being the second bright X-ray source discovered in that cluster.
It is peculiar for being the only burster where a smooth transition between bursts and mHz QPOs has been observed \citep{Linares2012}, and for having an unusually slow spin and high NS surface magnetic field strength \citep[$2 \times \lbrack 10^8 - 10^{10} \rbrack$~G;][]{papitto2011}.

We include its $t_{\rm rec}$ and $\dot m$ from Table 1 in \citet{Linares2012}, based on 46 {\it RXTE}-PCA observations of its 2010 outburst (the only one observed so far).
They used $c_{\rm bol}=1.132$ to convert from $L_X$ in the PCA band (2.5-25 keV) to  ``bolometric'' (0.01-50 keV) $L_{\rm acc}$. T5X2 showed very short $t_{\rm rec}$ down to about 5~min when the inferred $\dot{m}$ was high ($L_{\rm acc}=$18-50~\% of $L_{\rm Edd}$). At such high accretion rates, most other bursters rarely burst \citep{Linares2012,Cornelisse03}.

\subsection*{Circinus~X-1}

Since its discovery in the early days of X-ray astronomy \citep{Margon71}, the NS-LMXB Circinus~X-1 (Cir~X-1) has shown large variations in $L_X$ \citep[e.g.,][]{Armstrong13}.
It is a peculiar LMXB in a 16.6~d eccentric orbit, which showed XRBs again in 2010, after 25 years without bursting detected \citep{Linares2010}.

For Cir X-1, we list $t_{\rm rec}$ for the 4 PCA bursts reported by \citet{Linares2010} with uninterrupted observations from the previous burst (i.e., where $t_{\rm rec}$ could be measured unambiguously).
We use $L_X$ for those exact 4 bursts from \citet[][0.5-50~keV]{Linares2010}, and derive a $c_{bol}=1.018$ to convert $L_X$ to ``bolometric'' $L_{\rm acc}$ in the $0.1-100$ keV band.
The corresponding $\dot{m}$ is about 20\% of $\dot{m}_{\rm Edd}$.

\subsection*{4U~0614+09}

The persistent low-luminosity burster 4U~0614+09 features i) a tentative $P_{\rm orb}=0.86$~h \citep{Shahbaz08} which classifies it as a candidate UCXB, and ii) a lack of He lines in the optical spectrum \citep{Nelemans04}, which suggests the companion is a C/O WD.
This is puzzling since its burst behaviour seems consistent with pure He bursts \citep[see][and references therein]{Kuulkers10}.

For this burster we list the $t_{\rm rec}=12\pm3$~d measured by \citet{Linares20124U} using 1 year (2010-2011) of {\it Fermi} Gamma-Ray Burst Monitor (GBM) observations.
We also use their $L_{\rm acc}=[2.3\pm0.9]\times 10^{36}$~erg~s$^{-1}$, measured from {\it RXTE}-ASM and {\it Swift} Burst Alert Telescope (BAT) monitoring observations.
The updated 2010-2013 GBM results \citep{Jenke2016} and previous work by \cite{Kuulkers10} showed similar behavior: an $\dot{m}$ quite stable around 1\% $\dot{m}_{\rm Edd}$ and a $t_{\rm rec}$ of about two weeks.

\subsection*{2S~0918-549}

The persistent low-luminosity burster 2S~0918-549 also lacks He lines in the optical spectrum \citep{Nelemans04}, and the companion is thought to be an ONe WD \citep{Juett01}.
It has been classified as a UCXB based on a low optical/X-ray flux ratio \cite{Juett01} and on the likely $P_{\rm orb}= 17.1$~min \citep{zhong2011}.

We take the $t_{\rm rec}=56\pm12$~d measured by \citet{Jenke2016} from 2010-2013 GBM data, where they detected 10 XRBs from this source.
We also list the median and standard deviation of the 7 ``bolometric'' $L_{\rm acc}$ values reported by \citet{galloway2020}, which corresponds to 0.5$\pm$0.2\% of $L_{\rm Edd}$.

\subsection*{SRGA~J144459.2-604207}

The newly discovered burster and transient SRGA~J144459.2-604207 \citep{molkov2024}, was observed by the Imaging X-ray Polarimetry Explorer (IXPE) from February 27 to March 8 2024 \citep{papitto2024,Takeda2024}, during its outburst phase. From IXPE data, 52 XRBs were detected from this source \citep{papitto2024}. To include its $\dot m$, we convert the measured IXPE persistent count rate in the 2-8 keV band (from Figure 9 in \citealt{papitto2024}) to ``bolometric'' 0.1-100 keV $L_{\rm acc}$. Following \citet{papitto2024} and \citet{ng2024}, we used a power law model with photon index of 1.9 and a Galactic $N_{\rm H} = 2.9 \times 10^{22}~\rm{cm^{-2}}$. We thus list the obtained $L_{\rm acc}=5-22$ \% $L_{\rm Edd}$ (assuming $D=8.5$~kpc) and include the measured $t_{\rm rec}=2.2-6.2$~h.

\subsection*{Rapid~Burster}

The Rapid~Burster, MXB~1730-335, is a recurrent transient having multiple recorded previous outbursts \citep{masetti2002,heinke2024}. It has also shown different bursting epochs with appreciably different recurrence times, as well as type II bursts \citep{guerriero1999}.

Similarly to T5X2, \cite{Bagnoli2013} find that in the Rapid Burster, the bursting rate keeps increasing over a large range of persistent flux. To quantify their measurements, they convert from the PCA band (3-25 keV) to ``bolometric'' 0.1-200 keV $L_{\rm acc}$, using a bolometric correction factor which is not provided. We include the reported $t_{\rm rec}$ in Figure 3 and $\dot m$ \citep{Bagnoli2013}, using a distance of $D=7.9$~kpc. Burst recurrence times reach down to about $t_{\rm rec}=10$ min, while the accretion rate ranges around $L_{\rm acc}=9-40$ \% $L_{\rm Edd}$.

\subsection*{IGR~J17511-3057}

The burster and transient accreting millisecon pulsar IGR~J17511-3057 was observed by RXTE, INTEGRAL, Swift, as well as Chandra and XMM-Newton during its 2009 outburst \citep{Falanga2011}. The 13 XRBs were mostly discovered with RXTE (PCA) and INTEGRAL (JEM-X) observations. In order to estimate the persistent luminosity, \citet{Falanga2011} convert from the PCA band (2-60 keV) to ``bolometric'' 0.8-300 keV $L_{\rm acc}$. For the timing of XRBs, detections from Chandra and XMM-Newton were also used. We include the $t_{\rm rec}=7.1-15.8$ h and the corresponding $L_{\rm acc}=2-4$ \% $L_{\rm Edd}$, shown in Figure 10 of \citet{Falanga2011}, for a source distance $D=7$~kpc.

\subsection*{SAX~J1748.9–2021}

The atoll source, burster and transient SAX~J1748.9–2021 was observed during its 2015 outburst by INTEGRAL, XMM-Newton and Swift \citep{li2018}. Counting 56 total observed XRBs, 26 XRBs were detected by INTEGRAL in the hard state, 25 by XMM-Newton in the soft state, and 5 by Swift in both states. We include the $t_{\rm rec}-\dot m$ relation shown in Figure 4 of \citet{li2018}. Bursts have $t_{\rm rec}=0.9-2.1$~h and a corresponding range $L_{\rm acc}=13-27$~\% $L_{\rm Edd}$.

\subsection*{MAXI~J1816-195}

The transient MAXI~J1816-195 was discovered by MAXI during its only known outburst in 2022 (Negoro et al. 2022), and was found to be an accreting millisecond pulsar in a 4.8~h orbit (Bult et al. 2022), showing XRBs \citep{mandal2023}. Using Insight-HXMT
and NICER observations of 83 XRBs, \citet{wang2024} convert for the 17 joint XRBs the NICER persistent count rates (1-50 keV) to ``bolometric'' 0.1-200 keV $L_{\rm acc}$. To do so, they fitted the X-ray spectrum with a model consisting of an accretion disk with thermal Comptonization, assuming a Galactic $N_{\rm H}=2.4 \times 10^{-22}$~$\rm{cm^{-2}}$ and $D=6.5$~kpc. We include the $t_{\rm rec}-\dot m$ relation shown in Figure 7 of \citet{wang2024}, with $t_{\rm rec}=1.2-2$~h and $L_{\rm acc}=18-27$ \% $L_{\rm Edd}$.

\subsection*{Clocked~Burster}

For the transient atoll source known as the Clocked~Burster (GS~1826-24), \citet{Galloway2004} analyzed 24 XRBs observed by RXTE between 1997-2002, and used $c_{\rm bol}=1.68 \pm 0.02$ to convert from the PCA band (2.5-25 keV) to ``bolometric'' 0.1-200 keV $L_{\rm acc}$. We thus list the $t_{\rm rec}-\dot m$ relation shown in Figure 4 of \citet{Galloway2004}, featuring $t_{\rm rec}=3.5-5.9$~h and a corresponding $L_{\rm acc}=4-7$~\% $L_{\rm Edd}$. 

\subsection*{1RXS~J171824.2-402934}

The low-luminosity candidate UCXB 1RXS~J171824.2-402934 (the lowest-$\dot m$ NS in our sample) was discovered by \citet{kaptein2000}, with the detection of one XRB with the WFC.
It is one of the ``burst only'' or very-faint persistent NS-LMXBs, showing very low $L_X$ ($10^{34}-10^{35}$~erg~s$^{-1}$) when observed with sensitive pointed X-ray telescopes \citep{intZand09b,Armas13,Wijnands15}.
We include the values listed in Table 2 of \citet{intZand2007}, with $t_{\rm rec}=438-8254$~h and $L_{\rm acc}=0.03$~\% $L_{\rm Edd}$.
This $t_{\rm rec}$ is derived from the BeppoSAX-WFC database \citep{zand2004bepposax}, by dividing the total exposure time through the number of detected bursts.
A monitoring campaign by \citet{zand2009} detected a second XRB, putting the average waiting time between bursts at $t_{\rm wait}=125 \pm 88$~d (consistent with the $t_{\rm rec}$ range above).
While having very few XRBs detected, the burst characteristics and the low luminosity point to an extremely low $\dot m$ \citep{Zand2005}. 

\subsection*{1RXS~J180408.9-342058}
First detected by the ROSAT satellite \citep{Voges1999}, the burster 1RXS~J180408.9-342058 was observed during its 2015 outburst by Swift, INTEGRAL and NuStar \citep{Fiocchi2019}, showing spectral state transitions. \cite{Fiocchi2019} analyzed the quasi-simultaneous observations in the ``bolometric'' 0.8-200 keV energy band, using a bolometric correction factor $c_{\rm bol}=1.0$, assuming a distance of $D=5.8$ kpc. We include the values reported therein, with $t_{\rm rec}=1.1-2.2$ h and $L_{\rm acc}=7.2-10.4 \, \% \, L_{\rm Edd}$.

\begin{table*}
\center
\caption{Type I X-ray bursters shown in Figure~\ref{fig:NStrec}.}
\begin{minipage}{\textwidth}
\begin{tabular}{l c c c c c c c c c}
\hline\hline
\footnotetext{\textbf{Notes}: $^a$ 4U~0919-54; $^b$ GX~354-/+0; $^c$ Rapid~Burster; $^d$ IGR~J17480-2446, 11 Hz X-ray pulsar; $^e$ Clocked Burster, GS~1826-238; $^f$ value not reported, only unabsorbed bolometric flux provided; $^g$ Source type, same as \cite{galloway2020}; A = atoll source, C = UCXB (including candidates), D = “dipper,” E = eclipsing, G = globular cluster association, I = intermittent pulsar, M = microquasar, O = burst
oscillation, P = pulsar, R = radio-loud X-ray binary, S = superburst, T = transient, Z = Z-source.}
\footnotetext{\textbf{References}:(1)\cite{Galloway2008};(2)\cite{intZand2007};(3)\cite{Linares20124U};(4)\cite{Jenke2016};(5)\cite{papitto2024};(6)\cite{Linares2010};(7)\cite{Bagnoli2013};(8)\cite{Linares2012};(9)\cite{Falanga2011};(10)\cite{mandal2023};(11)\cite{Galloway2004};(12)\cite{li2018};(46)\cite{Fiocchi2019}.}
\footnotetext{$^1$\cite{fiocchi2011}; $^2$ \cite{Shahbaz08}; $^3$ \cite{intZand2007}; $^4$ \cite{zhong2011}; $^5$ \cite{intZand2007}; $^6$ \cite{intZand2007}; $^7$ \cite{ray2024}; $^8$ \cite{Kaluziensky1976}; $^9$ \cite{guver2021}; $^{10}$ \cite{mandal2023}; $^{11}$ \cite{cominsky1984}; $^{12}$ \cite{Galloway2008}; $^{13}$ \cite{revnivtsev2013}; $^{14}$ \cite{vincentelli2020}; $^{15}$ \cite{intZand2007}; $^{16}$ \cite{intZand2007}; $^{17}$ \cite{zolotukhin2011}; $^{18}$ \cite{papitto2011}; $^{19}$ \cite{sansom1993}; $^{20}$ \cite{riggio2011}; $^{21}$ \cite{chakrabarty1998}; $^{22}$ \cite{strohmayer2003}; $^{23}$ \cite{padilla2022}; $^{24}$ \cite{bult2022}; $^{25}$ \cite{stella1987}; $^{26}$ \cite{Mescheryakov2010}; $^{27}$ \cite{engel2012}; $^{28}$ \cite{gladstone2007}; $^{29}$ \cite{thorstensen1978}; $^{30}$ \cite{gladstone2007}; $^{31}$ \cite{prodan2015}; $^{32}$ \cite{cowley1979}; $^{33}$ \cite{altamirano2008}; $^{34}$ \cite{ZandBursters}.}
Burster & D & Type $^g$ & Spin Freq. $^{34}$ & $P_{\rm orb}$ & $L_{\rm acc}$ & $\dot{m}$ & $t_{\rm rec}$ & $c_{\rm bol}$ & Instrument \\
\hspace{0.15cm} Name & (kpc) & & (Hz) & (h) & $\left(\%\, L_{\rm Edd}\right)$ & $\left(10^2\,\rm{g\,cm^{-2}\,s^{-1}}\right)$ & (h) &  & (Refs.) \\
\hline
4U~0513-40 & 11.3 & CG & - & 0.3 $^{1}$ & 1.9 & $18.8 \pm 5.6$ & 35-63 & 2.9 & WFC\citepalias{intZand2007}\\
4U~0614+09 & 3 & ACRS & 415 & 0.8 $^{2}$ & 0.8-0.9 & 8-9 & 288$\pm$72 & 1.3-1.7 & GBM\citepalias{Linares20124U}\\
EXO~0748-676 & 6.3 & DEOT & 552 & 3.8 $^{3}$ & 1.0 & $9.9 \pm 3$ & 4.7-5.5 & 2.9 & WFC\citepalias{intZand2007}\\ 
1M~0836-425 & 8 & T & - & - & 33 & $327 \pm 98$ & 2.0 & $1.82 \pm 0.02$ & PCA\citepalias{Galloway2008}\\
2S~0918-549 $^a$ & 5 & C & - & 0.4 $^{4}$ & 0.3-0.7 & 3.3-6.4 & 1340$\pm$290 & 1.9 & GBM\citepalias{Jenke2016}\\
4U~1246-58 & 4.8 & C & - & - & 0.5 & $5 \pm 1.5$ & 139-417 & 2.9 & WFC\citepalias{intZand2007}\\
4U~1254-69 & 13.6 & DS & - & 3.9 $^{5}$ & 17.0 & $168 \pm 50$ & 36.1-53.7 & 2.9 & WFC\citepalias{intZand2007}\\
4U~1323-62 & 14 & D & - & 2.9 $^{6}$ & 4.3 & $42 \pm 13$ & 28-50 & 2.9 & WFC\citepalias{intZand2007}\\
SRGA~J144459.2-604207 & 8.5 & T & 448 & 5.2 $^{7}$ & 5-23 & 50-228 & 2.2-7.9 & - $^f$ & ART-XC\citepalias{papitto2024}\\
Cir~X-1 & 7.8 & ADMRT & - & 398 $^{8}$ & 17.6-22.4 & 174-221 & 0.33-0.5 & 1.02 & PCA-XRT\citepalias{Linares2010}\\
4U~1608-52 & 4.1 & AOST & 620 & 12.9 $^{9}$ & 5.2 & $50 \pm 15$ & 3.5 & $1.77 \pm 0.04$ & PCA\citepalias{Galloway2008}\\
4U~1636-53 & 5.3 & AOS & 581 & 3.8 $^{10}$ & 10.9 & $108 \pm 32$ & 7.9-9.9 & 2.9 & WFC\citepalias{intZand2007}\\
MXB~1659-298 & 10 & DEOT & 567 & 7.1 $^{11}$ & 4 & $40 \pm 12$ & 1.8 & 1.38 & PCA\citepalias{Galloway2008}\\
4U~1702-429 & 5.1 & AO & 329 & - & 2.7 & $27 \pm 8$ & 10.4-12.4 & 2.9 & WFC\citepalias{intZand2007}\\
1RXS~J170854.4-321857 & 11.7 & C & - & - & 1.5 & $15 \pm 4.5$ & 101-1904 & 2.9 & WFC\citepalias{intZand2007}\\
4U~1705-440 & 5.8 & AR & - & - & 7.2 & $133 \pm 40$ & 14.6-18.4 & 2.9 & WFC\citepalias{intZand2007}\\
XTE~J1710-281 & 16 & DET & - & 3.9 $^{12}$ & 0.8 & $7.9 \pm 2.4$ & 3.3 & $1.42 \pm 0.13$ & PCA\citepalias{Galloway2008}\\
SAX~J1712.6-3739 & 6.4 & CT & - & 1.0 $^{13}$ & 0.5 & $5 \pm 1.5$ & 345-6507 & 2.9 & WFC\citepalias{intZand2007}\\
1RXS~J171824.2-402934  & 7.3 & C & - & - & 0.03 & $0.3 \pm 0.09$ & 438-8254 & 2.9 & WFC\citepalias{intZand2007}\\
4U~1724-307 & 5.4 & ACG & - & - & 1.8 & $17.8 \pm 5.3$ & 45-69 & 2.9 & WFC\citepalias{intZand2007}\\
4U~1728-34 $^b$ & 4.2 & AOR & 363 & 2 $^{14}$ & 4.4 & $42 \pm 12.6$ & 3.0-3.4 & 2.9 & WFC\citepalias{intZand2007}\\
MXB~1730-335 $^c$ & 7.9 & DGRT & - & - & 9-40 & 89-395 & 0.17-1.10 & - $^f$ & PCA\citepalias{Bagnoli2013}\\
KS~1731-260 & 7.2 & OST & 524 & - & 5 & $49 \pm 14.7$ & 2.5-6.4 & $1.58 \pm 0.04$ & PCA\citepalias{Galloway2008}\\
SLX~1735-269 & 6 & CS & - & - & 1.0 & $9.9 \pm 3$ & 387-7301 & 2.9 & WFC\citepalias{intZand2007}\\
4U~1735-44 & 7.7 & ARS & - & 4.7 $^{15}$ & 28.6 & $282 \pm 85$ & 25.1-34.7 & 2.9 & WFC\citepalias{intZand2007}\\
SLX~1737-282 & 5.9 & C & - & - & 0.4 & $4 \pm 1.2$ & 412-7778 & 2.9 & WFC\citepalias{intZand2007}\\
1A~1742-294 & 7.2 & - & - & - & 2.7 & $27 \pm 8.1$ & 5.7-6.5 & 2.9 & WFC\citepalias{intZand2007}\\
SLX~1744-300 & 10.5 & T & - & 65 $^{16}$ & 3.2 & $32 \pm 9.6$ & 18-31.4 & 2.9 & WFC\citepalias{intZand2007}\\
SLX~1744-299 & 7 & T & - & 24 $^{17}$ & 1.9 & $19 \pm 5.7$ & 188-793 & 2.9 & WFC\citepalias{intZand2007}\\
GX~3+1 & 5.5 & AS & - & - & 22.2 & $219 \pm 66$ & 18.7-24.1 & 2.9 & WFC\citepalias{intZand2007}\\
T5X2 $^d$ & 6.3 & GOPT & 11 & 21.3 $^{18}$ & 18-50 & 178-495 & 0.08-60 & 1.13 & PCA\citepalias{Linares2012}\\
EXO~1745-248 & 5.5 & DGST & - & - & 5 & $50 \pm 15$ & 2.9 & $1.53 \pm 0.02$ & PCA\citepalias{Galloway2008}\\
SAX~J1748.9–2021 & 8.2 & AGIT & 410 & 8.7 $^{33}$ & 13-27 & 129-272 & 0.9-2.1 & - $^f$ & JEMX-XRT\citepalias{li2018}\\
4U~1746-37 & 11 & ADG & - & 5.7 $^{19}$ & 15 & $148 \pm 44$ & 1.0 & $1.45 \pm 0.05$ & PCA\citepalias{Galloway2008}\\
IGR~J17511-3057 & 7 & OPT & 245 & 3.5 $^{20}$ & 2-4 & 20-40 & 7.1-15.8 & - $^f$ & JEMX-PCA\citepalias{Falanga2011}\\
1RXS~J180408.9-342058 & 5.8 & - & - & - & 7.2-10.4 & 71-103 & 1.1-2.2 & 1.0 & XRT FPMA/B\citepalias{Fiocchi2019}\\
SAX~J1808.4-3658 & 3.6 & OPRT & 401 & 2.0 $^{21}$ & 1.1 & $10.9 \pm 3.3$ & 21 & $2.14 \pm 0.03$ & PCA\citepalias{Galloway2008}\\
XTE~J1814-338 & 8 & OPT & 314 & 4.3 $^{22}$ & 4 & $40 \pm 12$ & 1.7-6 & $1.86 \pm 0.3$ & PCA\citepalias{Galloway2008}\\
4U~1812-12 & 3.5 & AC & - & 1.9 $^{23}$ & 0.5 & $5 \pm 1.5$ & 61.3-99.1 & 2.9 & WFC\citepalias{intZand2007}\\
GX~17+2 & 12 & RSZ & - & - & 263 & $2587 \pm 776$ & 76-134 & 2.9 & WFC\citepalias{intZand2007}\\
MAXI~J1816-195 & 6.5 & T & - & 4.8 $^{24}$ & 18-27 & 178-267 & 1.2-2 & - $^f$ & FPMA\citepalias{wang2024}\\
4U~1820-303 & 6.1 & ACGRS & - & 0.2 $^{25}$ & 24.7 & $243 \pm 73$ & 22.8-30.4 & 2.9 & WFC\citepalias{intZand2007}\\
GS~1826-24 $^e$ & 6 & T & - & 2.1 $^{26}$ & 4-7 & 40-70 & 3.5-5.9 & $1.68 \pm 0.02$ & PCA\citepalias{Galloway2004}\\
XB~1832-330 & 8.3 & G & - & 2.2 $^{27}$ & 1.2 & $11.9 \pm 3.6$ & 20.4-35.2 & 2.9 & WFC\citepalias{intZand2007}\\
Ser~X-1 & 8.4 & ARS & - & - & 42.7 & $422 \pm 127$ & 46-104 & 2.9 & WFC\citepalias{intZand2007}\\
4U~1850-08 & 5.9 & ACGR & - & 0.3 $^{28}$ & 0.5 & $5 \pm 1.5$ & 1584 & 2.9 & WFC\citepalias{intZand2007}\\
Aql~X-1 & 5 & ADIORT & 549 & 19 $^{29}$ & 5.5 & $50 \pm 15$ & 3.5 & $1.65 \pm 0.05$ & PCA\citepalias{Galloway2008}\\
4U~1915-05 & 5.7 & ACD & 270 & 0.8 $^{30}$ & 1.1 & $11 \pm 3.3$ & 20-42 & 2.9 & WFC\citepalias{intZand2007}\\
M15~X-2 & 7.4 & CGR & - & 0.4 $^{31}$ & 1.2 & $12 \pm 3.6$ & 37-984 & 2.9 & WFC\citepalias{intZand2007}\\
Cyg~X-2 & 11.6 & RZ & - & 236 $^{32}$ & 105 & $1040 \pm 312$ & 1.0 & 1.38 & PCA\citepalias{Galloway2008}\\
\hline\hline
\end{tabular}
\end{minipage}
\label{tab:NS}
\end{table*}

\subsection{mHz QPOs: $t_{\rm mQPO} - \dot{m}$ correlation}
\label{sec:mhz}

Furthermore, we compile the timescale of the inverse QPO frequency defined as $t_{\rm mQPO} = \nu_{\rm mQPO}^{-1}$, for the bursters that have also shown mHz QPOs, as well as the $\dot{m}$ where these have been observed, listed in Table~\ref{tab:QPOs}. We include the errors on the mHz QPO frequency, and the range of $\dot m$ where applicable.

The frequency range of the observed mHz QPOs is $2.8-10$~mHz, which corresponds to timescales of $100-357$ s.
We find 11 sources with mHz QPO detections in the literature, shown in Table~\ref{tab:QPOs}. Only 8 of these are listed as bursters in Table~\ref{tab:NS}, since the remaining 3 do not fullfill our selection criteria (Section~\ref{sec:NStrec})\footnote{We find no reliable $t_{\rm rec}$ measurements for XTE~J1701-462 (the first known transient Z source), 4U~1730-22 and Swift~J1858.6–0814, but we report the measured $t_{\rm mQPO}$ and the associated $\dot{m}$.}.

In Figure~\ref{fig:QPOs}, we present the ranges of $L_{\rm acc}$, where mHz QPOs have been observed. The atoll sources cluster around $2-6 \%\, L_{\rm Edd}$, an order of magnitude below T5X2 and XTE~J1701-462, which are observed around $20-60\%\, L_{Edd}$. The general trend seen in Figure~\ref{fig:QPOs} shows 
longer timescales $t_{\rm mQPO}$ (lower $\nu_{\rm mQPO}$)
for mHz QPOs at higher accretion rates.
Thus, we find that our sample of 22 observations of mHz QPO timescale and $\dot{m}$ show a negative correlation, with a Pearson correlation coefficient $c_{\rm pearson}=-0.8546$ (for $P_{\rm value}=4\times10^{-7}$).

T5X2 is the only burster (and mHz QPO source) with a known spin frequency below 200~Hz, and it showed mHz QPOs at high $\dot{m}$ and high $t_{\rm mQPO}$ \citep[][see Tables \ref{tab:NS} and \ref{tab:QPOs}]{Linares2012}.
Interestingly, we find that XTE~J1701-462 occupies the exact same  $t_{\rm mQPO}-\dot{m}$ region (Figure~\ref{fig:QPOs}).
This suggests that XTE~J1701-462 might be a slow rotator, if the mHz QPO frequency and mass accretion rate are indeed dependent on the NS spin frequency.
Alternatively, we could be witnessing marginally stable burning of different elements (hydrogen and helium) at different mass accretion rates (a few percent and a few tens of percent, respectively).

\begin{table*}
	\caption{Bursters with detected mHz QPOs.}
    \begin{minipage}{\textwidth}
    \centering
	\begin{tabular}{l c c c c c c} 
		\hline
  \footnotetext{\textbf{References}:(8)\cite{Linares2012};(12)\cite{tse2023};(13)\cite{revnivtsev2001};(14)\cite{strohmayer2011};(15)\cite{mancuso2019};(16)\cite{mancuso2023};(17)\cite{strohmayer2018};(18)\cite{tse2021};(19)\cite{buisson2020};(43)\cite{molkov2005}.}
		Burster & Spin Freq. (Hz) & D (kpc) & Date (MJD) & QPOs frequency (mHz) & $\dot m(\rm{10^2\, g\,cm^{-2}\,s^{-1}})$ & References\\
		\hline
		XTE~J1701-462 & - & 6.4 & 53817 & 3.5 & 510 & \citepalias{tse2023}\\
		     & & & 53859 & 4.8 & 320 & \\
		     & & & 53996 & 5.6 & 290 & \\
           & & & 54279 & 3.5 & 190 & \\
        T5X2 & 11 & 6.3 & 55485.46 & $2.8 \pm 0.2$ & 450 & \citepalias{Linares2012}\\
             & & & 55485.63 & $3 \pm 0.5$ & 490 & \\
             & & & 55487.43 & $4.2 \pm 0.2$ & 560 & \\
             & & & 55487.62 & $4.2 \pm 0.2$ & 600 & \\
             & & & 55488.26 & $4 \pm 0.5$ & 590 & \\
             & & & 55489.59 & $3.8 \pm 0.2$ & 480 & \\
             & & & 55490.63 & $2.9 \pm 0.2$ & 450 & \\
        4U~1608-52 & 620 & 4 & 50145-50148 & $7.5 \pm 0.2$ & 50-80 & \citepalias{revnivtsev2001} \\
                   & & & 50896-50899 & $7.6 \pm 0.3$ & 40-60 & \\
        4U~1636-536 & 581 & 5 & 50174-51210 & $8.8 \pm 0.1$ & 30 & \citepalias{revnivtsev2001} \\   
                   & & & 46285 & $7.8 \pm 0.2$ & 30-50 &  \\
        4U~1323-619 & - & 14 & 55648 & 8.1 & 2\footnote{$\dot m$ inferred from luminosity reported in \cite{tse2021}.} & \citepalias{strohmayer2011} \\
        EXO~0748-676 & 552 & 7.1 & 53071-54439 & 5-13 & 10-30 & \citepalias{mancuso2019} \\
        4U~1730-22 & 585 & 6.9 &59624-59809 & 4.5-8.1 & 10-40 & \citepalias{mancuso2023} \\
        GS~1826-238 & - & 5.7 & 58005-58011 & 8 & 150 & \citepalias{strohmayer2018} \\
        1RXS~J180408.9-342058 & - & 5.8 & 57087 & 5-8 & 3 & \citepalias{tse2021} \\  
        Swift~J1858.6–0814 & - & 12.8 & 58930 & $9.6 \pm 0.5$ & 20 & \citepalias{buisson2020} \\
        SLX~1735-269     & - & 8.5 & 52902 & 10 & 20 & \citepalias{molkov2005} \\
		\hline
	\end{tabular}
    \end{minipage}
    \label{tab:QPOs}
\end{table*}

\subsection{Results: neutron star burst recurrence times}
\label{sec:NStrecRES}

\begin{figure*}
    \includegraphics[width=1.9\columnwidth]{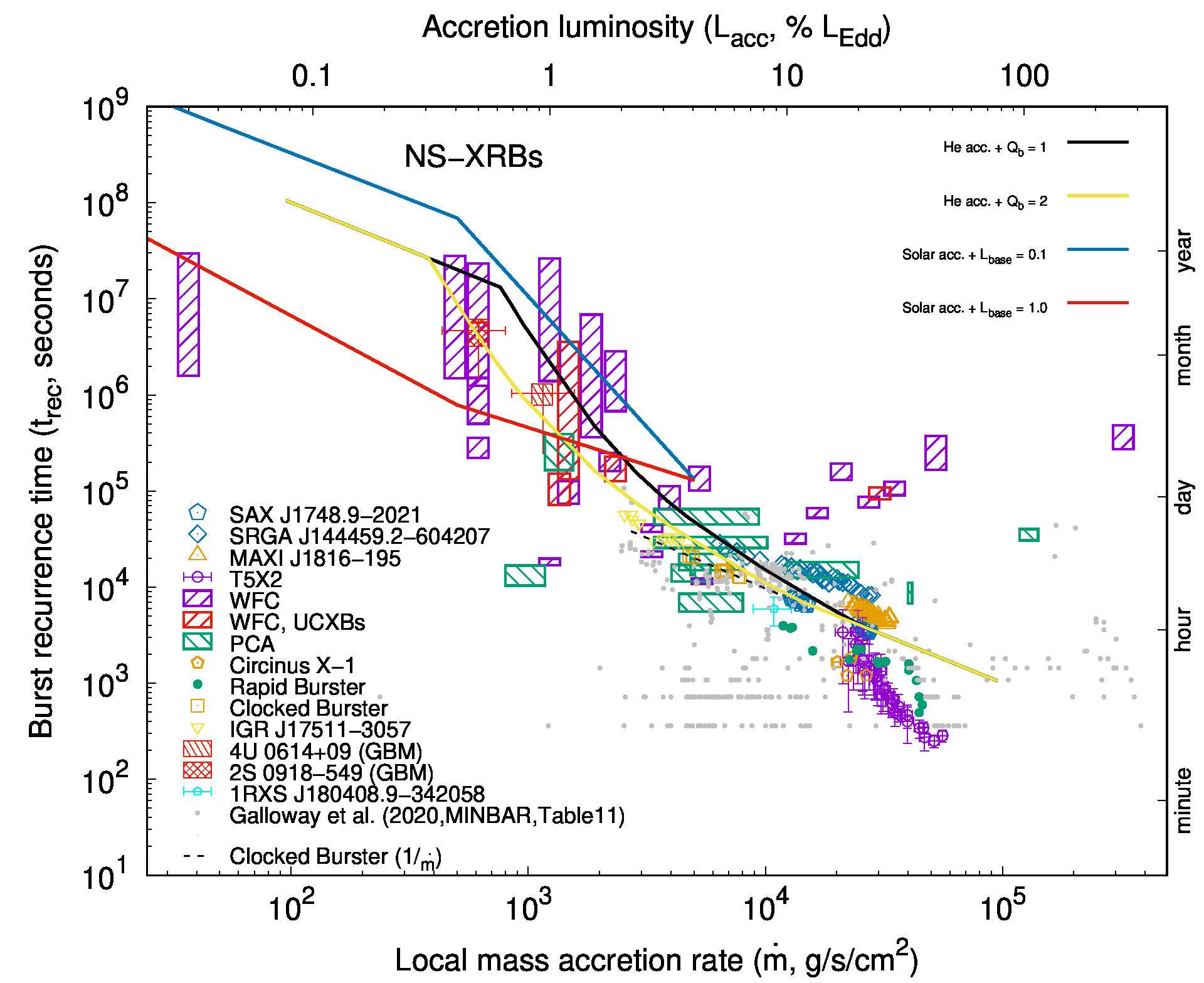}
    \caption{Observations and models of $t_{\rm rec}$ of type I X-ray bursts on NSs listed in Table~\ref{tab:NS}, as a function of the persistent luminosity and corresponding $\dot m$. Compilation made using measurements from BeppoSAX-WFC (\protect\citealt{intZand2007}), RXTE-PCA (\protect\citealt{Galloway2008}), the MINBAR catalog \citep{galloway2020} and highlighted sources, as indicated. The UCXB systems are: 4U~0614+09 (at $0.01\, L_{\rm Edd}$), 2S~0918-549 ($0.003-0.07\,  L_{\rm Edd}$, coinciding in position with 4U~1850-08), and all other red rectangles shown. Pure He accretion models (\textsc{SETTLE}, see Section~\ref{sec:SETTLE}) with different $Q_{\rm b}$ are shown ($\rm MeV\,nucleon^{-1}$, black and yellow lines). The blue and red lines denote models with solar composition accretion, computed using \textsc{SHIVA} (see Section~\ref{sec:SHIVA}).}
    \label{fig:NStrec}
\end{figure*}

\begin{figure*}
	\includegraphics[width=1.5\columnwidth]{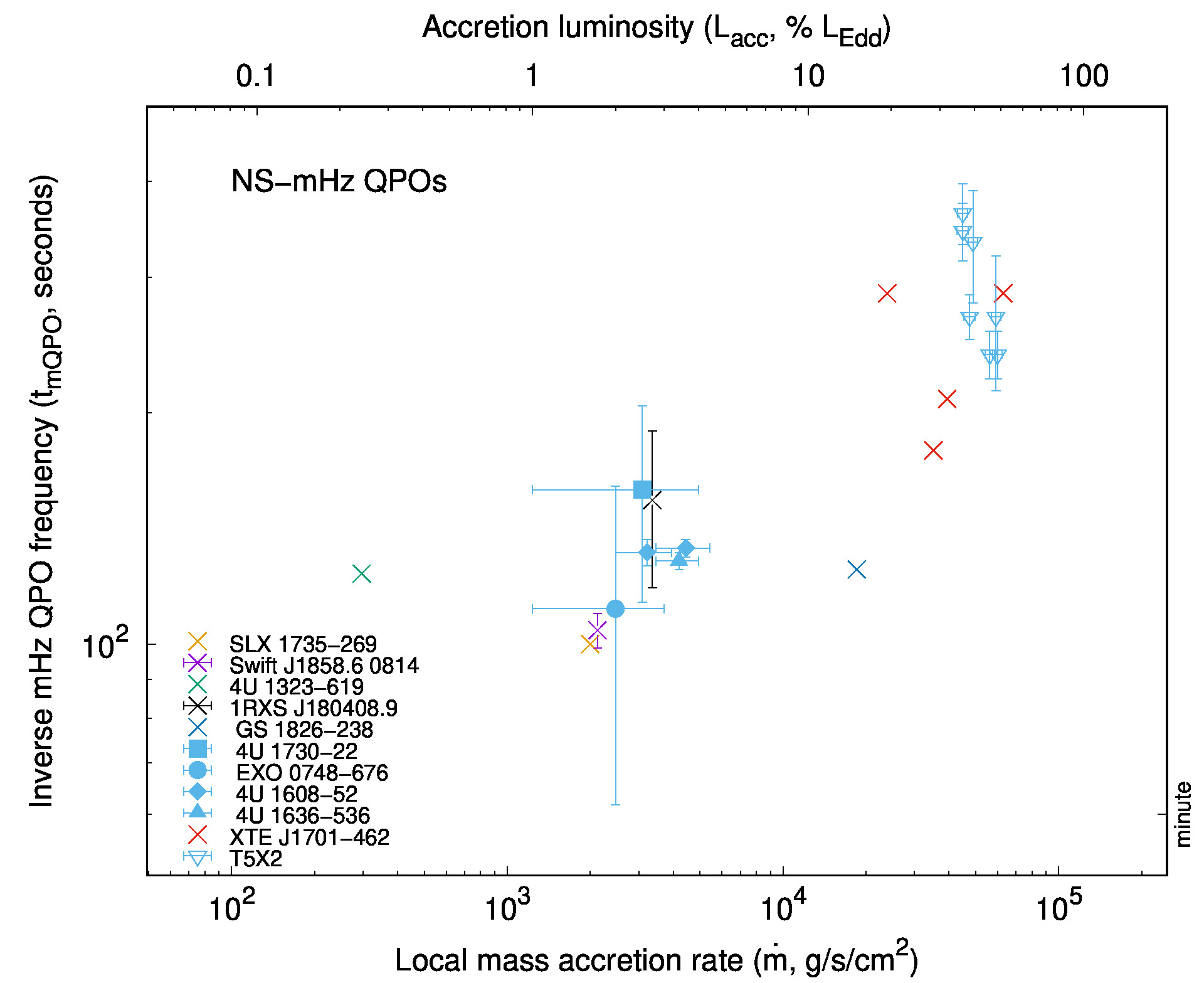}
    \caption{The inverse frequency of mHz QPOs listed in Table~\ref{tab:QPOs}, as a function $L_{\rm acc}$ and $\dot m$. Filled symbols refer to known fast rotators ($f_{\rm spin}=(200-600)$ Hz), empty symbols to known slow rotators ($f_{\rm spin}<200$ Hz), and the x symbols (with different colors, and errors where indicated) to bursters with unknown Spin Freq.}
    \label{fig:QPOs}
\end{figure*}

We show in Figure~\ref{fig:NStrec} the observed $t_{\rm rec}$ vs $\dot m$ in XRBs, over a range of 5 orders of magnitude in $t_{\rm rec}$ and 4 orders of magnitude in $\dot m$. The WFC measurements from \cite{intZand2007} are separated in two categories, the purple rectangles denoting LMXBs with $P_{orb}\geq 1$ h or unknown period, and the red rectangles (including the GBM measurements of 4U 0614+09 and 2S 0918-549) showing UCXBs. The 7 UCXBs reported in Table~\ref{tab:NS} are the systems: 4U~0513-40, 4U~0614+09, 2S~0918-549, 4U~1820-303, 4U~1915-05, 4U~1850-08 and M15~X-2. Most UCXBs are observed accreting around 1\% of $L_{\rm Edd}$, with the exception of 4U~1820-303, which is at $L_{\rm acc}=25$ \% $L_{\rm Edd}$.

The green rectangles indicate PCA measurements from \cite{Galloway2008}. The remaining sources, namely SAX~J1748.9-2021, SRGA~J144459.2-604207, MAXI~J1816-195, T5X2, Cir~X-1, Rapid~Burster, GS~1826-24 (Clocked~burster) and IGR~J17511-3057, are shown separately with different symbols. We also show the $1/\dot m$ relation ($t_{\rm rec}\propto\dot m^a$, with $a=-1.05\pm 0.02$) for the bursting evolution of the Clocked~burster (dotted black line in Figure~\ref{fig:NStrec}, fit from Figure 4 of \citealt{Galloway2004}). The source which follows a similar evolution in the $t_{\rm rec}\propto\dot m^a$ is SAX~J1748.9-2021, with $a=-1.02 \pm 0.03$ \citep{li2018}.

For $ L_{\rm acc} > 10\%\, L_{\rm Edd}$, a bifurcation in the frequency of observed XRBs is seen. Some systems burst more frequently for higher accretion rates \citep{Linares2012}, whereas others show less frequent bursts \citep{Cornelisse03}. 
The mechanism behind this observed difference may be attributed to 
a changing accretion geometry \citep{bildsten1998}, to the ignition location with respect to the pole of the NS, or to the spin rate \citep{Linares2012,Cavecchi2017}. Mixing of the ashes with burning layers has also been invoked as a possible explanation \citep{Galloway2018}.
In the pure He ignition regime, a $t_{\rm rec}\propto\dot m^3$ relation is expected \citep{cumming2006}.
\cite{Linares2012} found $a=-3.2\pm 0.5$ for the lower $\dot m$ bursts from T5X2 (their regime B).
In summary, XRBs at high accretion rates ($ \dot m > 10\%\, \dot{m}_{\rm Edd}$) show a complex phenomenology, deviations from a linear $t_{\rm rec}-\dot m$ relation and evidence for a possible influence of the underlying NS spin \citep{Linares2012,Cavecchi2017, Galloway2018}.

At low $\dot m$, the overall XRB trend is roughly consistent with a $1/\dot m$ relation (see Section~\ref{sec:Discussion} for further discussion).
Two sources clearly depart from such overall trend: EXO~0748$-$676 and XTE~J1710$-$281 ($\dot m \simeq 1\%~\dot{m}_{\rm Edd}$ and $t_{\rm rec}$ of a few h).
They are both "dippers" and eclipsing systems, which suggests that the mass accretion rates may be underestimated (i.e., that inclination effects are critical and the assumption of isotropic accretion luminosity may break).

\section{White dwarfs}
\label{sec:WDtrec}
 
In this section we compile a list of recurrence times of known recurrent novae (RNe), together with $\dot m$, estimated from the average $\dot M$ provided from observational constraints (explained separately for each source listed in Table~\ref{tab:WD}). These RNe can be identified by having at least two confirmed observed eruptions, or alternatively based on the presence of nova super-shells, that confirm past explosions (\citealt{darnley2019recurrent}).

\subsection{The RNe sample: recurrence times and accretion rates}

Our list consists of the Galactic RNe with measured recurrence times  \citep{Schaefer_2010,Schaefer2022}, as well as confirmed RNe in the Large Magellanic Cloud (LMC, from \citealt{rosenbush2020}) and M31 \citep{shafter2015}. We only list RNe which have empirical constraints on both $t_{\rm rec}$ and $\dot M$. This means that the RNe V394 CrA and V745 Sco listed in \cite{Schaefer_2010} (with respective $t_{\rm rec}$ of 30 and 23 yr) are not shown in Table~\ref{tab:WD}, because to the best of our knowledge no observational constraints exist on the $\dot M$ of these systems (only model estimations on the $M_{\rm WD}$ and the $\dot M$ can be found in \citealt{Shara_2018}).

Calculating $\dot m$ will allow us to compare XRBs from accreting NS with RNe from accreting WDs. Using the known relation for the WD radius \citep{chandrasekhar1933}:

\begin{equation}
\label{eq:wdmr}
R_{\rm WD}\,(10^8\,\rm {cm})=M_{\rm WD}^{-1/3}\,(\rm M_{\odot}),
\end{equation} \\
and the $M_{\rm WD}$ tabulated for each system, we calculate $\dot m = \dot M /4\pi R^2_{\rm WD}$, and report them in Table~\ref{tab:WD}. This is a simplified mass-radius relation for massive WDs, but sufficient for our goal of estimating $\dot m$. It is also useful to mention that $M_{\rm WD}$ is high in many RNe (see also \cite{Shara_2018}, where the 10 Galactic RNe analyzed, have an average $M_{\rm WD}$ of 1.31$M_{\rm \odot}$).

The eruption $t_{\rm rec}$ for 9 out of the 11 known Galactic RNe are reported in \cite{Schaefer_2010}, where $t_{\rm rec}$ is calculated taking possible missed eruptions into account. This is done by calculating statistically the efficiency of the discovery of a nova event, as well as studying the time interval ranges between eruptions. The justification for this assumption is based on the fact that a multitude of reasons can lead the observing runs to miss a nova event \citep{Schaefer_2010}. The $t_{\rm rec}$ and $\dot m$ of the 12 RNe are listed in Table~\ref{tab:WD}. We present in Figure~\ref{fig:WDtrec} the $t_{\rm rec}$ vs $\dot m$ relation.

CNe are expected to have $t_{\rm rec}$ that are much longer than 100 yr, and so much lower $\dot M$, and possibly lower $M_{\rm WD}$ are expected to be present in these systems, in general. These estimates have been reported in multiple works (see e.g., \citealt{Shara_2018}). We now move on to explain how the $\dot M$ listed in Table~\ref{tab:WD} are estimated for each RN.

\subsection*{T~Pyx}
One of the most well known, and the first to be classified as RN, T~Pyxidis first nova eruption was in 1890, discovered by H.Leavitt \citep{pickering1913}. Since then it has had 5 confirmed eruptions, the last one being in 2011. \cite{Selvelli2008} estimate $\dot M$ from the International Ultraviolet Explorer (IUE) spectroscopic and optical data, by assuming that the accretion disk is heated by viscous dissipation of gravitational energy. From their Equation (10), given the luminosity of the disk and a range of masses $M_{\rm WD}=1.00-1.40$~$\rm M_{\odot}$ , they report a range of $\dot M=(0.84-3.66)\times 10^{-8}\, \rm M_{\odot}\, yr^{-1}$. They also estimated $\dot{M}$ from the boundary layer luminosity, yielding a value of $\dot M=1.6\times 10^{-8}\,\rm M_{\odot}\, yr^{-1}$. The third estimate used was the absolute optical magnitude versus the accretion rate for the “reference” alpha-disc model of \cite{shakura1973} ($M_v-\dot M$), and depending on the inclination, one can get values for $\dot M$ as high as $\dot M=2.85\times 10^{-8}\, \rm M_{\odot}\, yr^{-1}$ for T Pyx. The value we adopt here is the range of $\dot M$ given in Table 3 of \cite{Selvelli2008}, with $\dot M=(0.84-3.66)\times 10^{-8}\,\rm M_{\odot}\, yr^{-1}$, for a range of $M_{\rm WD}=1.00-1.40\,\rm M_{\odot}$ and $t_{\rm rec}=19$~yr \citep{Schaefer_2010}.

\subsection*{IM~Nor}
IM~Normae shows similarities to T~Pyx, such as the orbital light curves. It was discovered in 1920 \citep{1920Woods}, and since then, not much was known about the system until the second eruption in 2002 \citep{liller2002}. The recurrence time listed in Table~\ref{tab:WD} is an upper limit \citep{Patterson2022}, and the intrinsic value is given by $t_{\rm rec}=82/N$, with N being the uncertain number of missed eruptions (taking the values of 1, 2, 3 or 4). \cite{Patterson2022} use the eclipse times to derive $P_{orb}$ and its derivative $\dot P_{\rm orb}$ of the system, which they then combine with the masses of the binary system (assuming a WD mass $M_{\rm WD}=1.0\,\rm M_{\odot}$ and a companion mass of $0.2 \rm M_{\odot}$), to infer a mass transfer rate (assuming this is approximately equal to $\dot M$) of $\dot M=6.6\times 10^{-8}\,\rm M_{\odot}\, yr^{-1}$, which we list, together with $t_{\rm rec}\leq 82$~yr.

\subsection*{CI~Aql}
CI~Aql had its first confirmed eruption in 1917 \citep{Schaefer_2010}. \cite{Sion2019} obtained an estimate for $\dot M$ by fitting Far UV spectrum data (FUV, $1170 \textup{~\AA}$ to $1800 \textup{~\AA}$) from Hubble Space Telescope (HST), to model WD accretion disks and photospheres. Together with the Gaia parallax (giving a distance of 3.062 kpc), they constructed models of modified "standard disk" \citep{godon2017} of \cite{wade1998}, choosing the disk radius and photosphere temperature, as well as the system parameters (inclination, $M_{\rm WD}$, $\dot M$), in order to best fit the spectra obtained. They estimated a value of $\dot M=4\times 10^{-8}\rm M_{\odot} yr^{-1}$. Similarly to \cite{Selvelli2008}, using P and $\dot P$, \cite{Sion2019} are able to also estimate the WD $\dot M$ in the range $dM/dt=(4.8-7.8)\times 10^{-8}\,\rm M_{\odot}\, yr^{-1}$. We use and plot the value of $\dot M=4\times 10^{-8}\rm M_{\odot} yr^{-1}$, in conjunction with $t_{\rm rec}=24$~yr.

\subsection*{V2487~Oph}
V2487~Ophiuchi was first discovered on June 15 1998 \citep{lynch2000}. \cite{schaefer2022V2487} obtained and combined a large set of photometric observations, spanning from 2002 to 2018. By constructing the lightcurve and interpreting the shape of the spectral energy density (SED) profile, \cite{schaefer2022V2487} argue that the disk dominates the flux, and estimate $\dot M$ by fitting to the available SED an alpha-disc model \citep{shakura1973} plus a blackbody model (Section 2.4 therein). Another way they obtain $\dot M$ is through the absolute magnitude of the disc in the V-band. Here, using only the alpha-disc model, in the V-band the luminosity is almost entirely from the disc, and considering the possible range of inclination angles ($0^o$ to $30^o$), \cite{schaefer2022V2487} get a range of $\dot M=(8-57)\times 10^{-8}\,\rm M_{\odot}\, yr^{-1}$. The two methods give $\dot M=5\times 10^{-8}\,\rm M_{\odot}\, yr^{-1}$ and $\dot M=20\times 10^{-8}\,\rm M_{\odot}\, yr^{-1}$ respectively, with the central value being $\dot M=9\times 10^{-8}\,\rm M_{\odot}\, yr^{-1}$. In our Table~\ref{tab:WD} and Figure~\ref{fig:WDtrec}, we include the full range of estimated $\dot M=(8-57)\times 10^{-8}\rm M_{\odot} yr^{-1}$, with $t_{\rm rec}=18$~yr.

\subsection*{U~Sco}
U~Scorpii had its first identified eruption in 1863 \citep{Schaefer_2010}. Following its 2010 outburst, \cite{maxwell2013} aimed to study the system in quiescence, by obtaining Very Large Telescope (VLT) and Southern African Large Telescope (SALT) spectra 18 and 30 months after the outburst respectively. A model of the accretion disc and the companion star as a single stellar model was fitted to the VLT spectra. This method gave low values of $\dot M$, in the range $\dot M=(4-7)\times 10^{-9}\,\rm M_{\odot}\, yr^{-1}$. The second method used, common for RNe, is the strength of the He II $4686 \textup{~\AA}$ line, which is shown to serve as an empirical relation of $\dot M$ in compact CVs \citep{patterson1985}. The latter gave a range of accretion rates $\dot M=(1-9)\times 10^{-7}\,\rm M_{\odot}\, yr^{-1}$. \cite{maxwell2013} argue that the discrepancy between the values is mainly due to different epochs during outbursts, since shortly after the outburst the disc hasn't reached its steady state, which took 18 months to fully reform. We take the $\dot M=(4-7)\times 10^{-9}\,\rm M_{\odot}\, yr^{-1}$ accretion rate, being the more representative measure of the average $\dot M$ between eruptions, for $t_{\rm rec}=10$~yr. 

\subsection*{T~CrB}
T~Coronae~Borealis is among the best known RNe, showing recorded eruptions in 1866 and 1946, but also possible eruptions in 1217 and 1787 \citep{schaefer2023}. With the next outburst predicted to occur imminently (see \cite{jose2025}, for new nucleosynthesis predictions
for the forthcoming outburst), monitoring campaigns and observations are being prepared to study the system in detail once again. To estimate $\dot M$, \cite{Zamanov2023} performed and analyzed UBV photometry during and after its superactive state between 2016 and 2023. From their analysis, they obtained the $(U-B)_0$ and $(B-V)_0$ colors of T CrB, and subtracted the contribution of the 'cool' component of the system (red giant). This left them with the hot component of the system (the WD), where using a distance of 914 pc and calculating the dereddened magnitude of the WD, they were able to calculate the effective radius and optical luminosity of the WD. Finally, by linking the optical luminosity to $\dot M$ through the relation \\
\begin{equation}
    L=0.5\,G\,M_{\rm wd}\, \dot M_{\rm \alpha}\, cos~i\, R^{-1}_{\rm wd},
\end{equation}
with $i$ being the inclination angle, assuming the main contributor to the optical luminosity is the accretion disc, \citet{Zamanov2023} obtained $\dot M$ in the range $\dot M=(0.27-4.05)\times 10^{-8}\,\rm M_{\odot}\, yr^{-1}$, which we use together with $t_{\rm rec}=80$~yr \citep{Schaefer_2010}.

\subsection*{RS~Oph}
\cite{nelson2011} obtained and analyzed light curves and spectra for RS~Oph in 2007 and 2008 using Chandra and XMM-Newton. It became apparent that the source was variable, suggesting that this was due to accretion-induced flickering. They estimated the amplitude of the flickering to constrain $\dot M$, following \cite{bruch1992}. This method gave lower limits on $\dot M$ in the $10^{-10}-10^{-9}\, M_{\odot}\, yr^{-1}$ range (not sufficient to explain such frequent eruptions). To place an upper limit on $\dot M$, a cooling flow model was used to fit the spectra \citep{mukai2003} and an optically-thick region of the boundary layer of the accretion disc was assumed \citep{patterson1985}. This called for the addition of a blackbody component to the model fit. Using values for the distance between 1.6 and 2.45 kpc, the cooling flow plus optically-thick disc can produce a value of $\dot M$ in the range $\dot M=(0.2-3.6)\times 10^{-8}\,\rm M_{\odot}\, yr^{-1}$, which we include together with $t_{\rm rec}=15$~yr.

\subsection*{V3890~Sgr}
The first eruption of V3890~Sgr occurred in 1962, and was published by \cite{dinerstein1973}. By analyzing SALT optical photometry and spectroscopy data of the system at quiescence, \cite{mikolajewska2021} deduce the $P_{\rm orb}$ of the binary and the mass of the WD (shown in Table~\ref{tab:WD}). Through the presence and intensity of the He II $4686 \textup{~\AA}$ line, they place a lower limit value on the accretion rate of $\dot M\geq 10^{-8}\,\rm M_{\odot}\, yr^{-1}$. Secondly, from the luminosity of the blue continuum of the source, as well as indicating that the supersoft X-ray emission begins $\sim$8.5 d later, and ends by day 26 of the nova outburst, they also infer an upper limit of $\dot M\leq10^{-7}\,\rm M_{\odot}\, yr^{-1}$. Thus, we use $\dot M = (1-10)\times 10^{-8}\,\rm M_{\odot}\, yr^{-1}$, and $t_{\rm rec}=22$~yr.

\subsection*{KT~Eri}
\cite{Schaefer2022} also used the alpha-disc model to convert the absolute optical magnitude of the system KT~Eridani into an estimate of $\dot M$. This was done by calculating the average V-magnitude of the system when in quiescence, and then correcting with the absolute optical magnitude of the disc alone. The latter can be calculated from fitting a blackbody plus alpha disc model to the SED profile of the system. The accretion luminosity gives out a flux both in the optical, as well as the ultraviolet, but \cite{Schaefer2022} argue that the $M_{\rm V_{\rm disc}}$ is a good proxy of $\dot M$. The value they get is $\dot M=3.5\times 10^{-7}\,\rm M_{\odot}\, yr^{-1}$, which is a very high rate, and is sitting in the zone of stable and continuous H burning on the surface of the WD (see also Figure 8 of \citealt{Schaefer2022}). 

A key difference in this system lies in the way $t_{\rm rec}$ is calculated. Instead of looking for observations of previous eruptions and using the intervening quiescent epoch, archival sky photographs worldwide were investigated to constrain the possible times of prior eruptions. This gave the only possible ranges of $t_{\rm rec}>82$~yr and $40 < t_{\rm rec} < 55$~yr. From the optical, X-ray light curves and the radial velocity curves, the mass of the WD in KT Eri was constrained within $M_{\rm WD}=1.25 \pm 0.03\,\rm M_{\odot}$. Thus, we include the aforementioned mass range and an accretion rate $\dot M=(0.83-8.2)\times 10^{-7}\,\rm M_{\odot}\, yr^{-1}$. Combining the archival constraints, and the average $\dot M$ and $M_{\rm WD}$ $t_{\rm rec}$ derivation, including uncertainties, we take the range of $t_{\rm rec}=5-50$~yr given in \citealt{Schaefer2022}. 

\subsection*{M31N~2008-12a}
M31N~2008-12a was first discovered during a nova eruption in 2008 \citep{nishiyama2008}. The curious case of this RN erupting every year \citep{darnley2016} requires model assumptions of (simultaneously) very high values of $M_{\rm WD}$, $\dot M$ and initial WD luminosity to power such frequent nova eruptions. To obtain $\dot M$ in quiescence from photometric observations of M31N from HST, \cite{darnley2017} follow up the 2015 eruption and construct the SED of the system. By constructing a model accretion disc to be fitted to the SED, they arrive at a range of $\dot M \sim 10^{-7}-10^{-5}\,\rm M_{\odot}\, yr^{-1}$. This is the highest accretion rate inferred to date in any RN. The $P_{\rm orb}$ of the system hasn't been determined, but constraints have been placed due to knowledge of the $M_{\rm WD}$ and the radius of the donor (see Sec. 5.3 in \citealt{darnley2017}), which we report in Table~\ref{tab:WD}.

\subsection*{Nova~LMC~1968}
Nova~LMC~1968 (in the Large Magellanic Cloud) was the first extragalactic nova discovered \citep{shore1991}. \cite{rosenbush2020} lists known and candidate extragalactic RNe, where Nova~LMC~2016-01a is reported as an extragalactic RN with 6 known eruptions (newest reported in \citealt{darnley2024}), and a $t_{\rm rec}=6.2$~yr. In that work, using an estimated mass ejection of $M_{\rm ej}=10^{-7.3\pm0.5}$ $\rm M_{\odot}$ from photoionization modeling, \cite{shore1991} inferred a lower limit of $\dot M\geq 10^{-8.6\pm0.5}$ $\rm M_{\odot}$ $yr^{-1}$ (also mentioned in \citealt{kuin2020}), which we include. From the observed UV luminosity, it is also argued that a He-enriched mixture for the fuel is more appropriate. 

\subsection*{Nova~LMC~1971b}
First discovered in 2009 \citep{liller2009} and also reported in \cite{rosenbush2020}, Nova~LMC~2009-2a has had two eruptions confirmed so far, the first one being Nova~LMC~1971b (cofirmed by position measurements within 0.11'', see \citealt{bode2016}) yielding $t_{\rm rec}=38$ yr. By obtaining optical and near-infrared photometry using the Small and Moderate Aperture Research Telescope System (SMARTS) 1.3 m telescope, as well as spectroscopy, \cite{bode2016} measure the intensity of the He II 4686 line, and also argue that from the shape of the SED, the luminosity of the system is dominated by the disc. By fitting a model disc to the data, they derive an accretion rate range of $\dot M=(1.1-8.3) \times 10^{-7}\,\rm M_{\odot}\, yr^{-1}$, which we list, as well as with $t_{\rm rec}=38$~yr.

\subsection{Extragalactic RNe without $\dot M$ constraints}
\label{subsec:extragalactic RNe}

Other extragalactic RNe with confirmed eruptions, as well as candidate RNe in the Andromeda galaxy (M31), Large Magellanic Cloud (LMC) and Small Magellanic Cloud (SMC) have been reported by \cite{shafter2015} and \cite{rosenbush2020} respectively (as well as telegrams and discovery reports therein). These systems either lack detailed photometry and spectroscopy for analysis, or have not been studied in detail as of yet. We only report the recurrence times found for each source below.

From the confirmed RNe in M31 \citep{shafter2015}, M31N~1926-07c (having detected subsequent eruptions 1997-10f and 2008-08b, Table 3 of \citealt{shafter2015}) and M31N~2017-01e (with all recorded eruptions being M31N~2012-01c, 2017-01e, 2019-09d, and 2022-03d, see \citealt{shafter2022}) have $t_{\rm rec}$ of 2.8 and 2.5 yr, respectively. One of the most recently identified and most luminous known RNe in M31 is M31N~2013-10c (reported in \citealt{shafter2023, shafter2024}), having an upper limit of $t_{\rm rec}=10.1$ yr, and being the $21^{st}$ RNe identified so far in M31.

\subsection{Results: white dwarf eruption recurrence times}

\begin{figure*}
	\includegraphics[width=1.5\columnwidth]{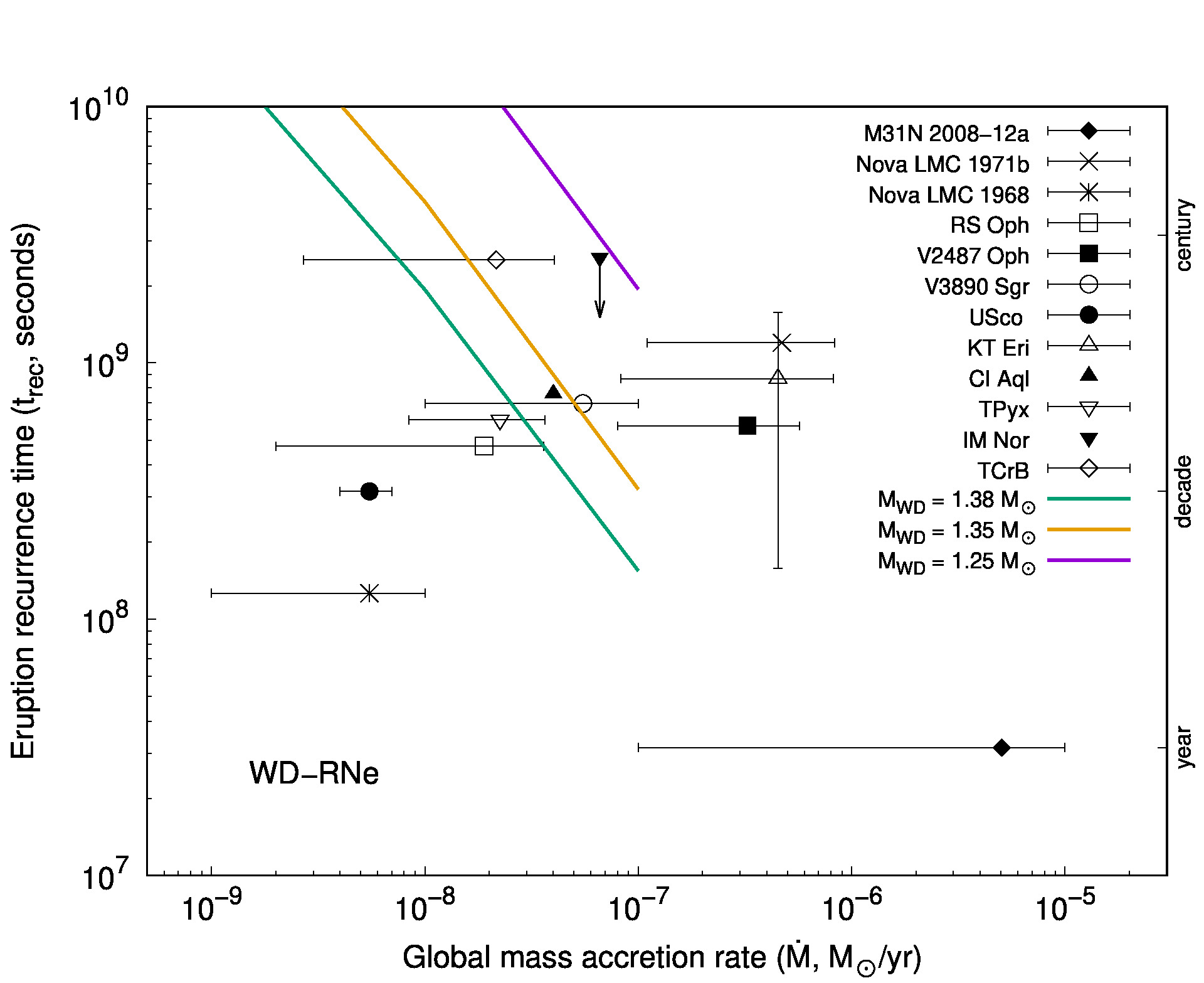}
    \caption{Recurrence times of nova eruptions on RNe listed in Table~\ref{tab:WD}, as a function of $\dot M$. Lines denote theoretical models of $t_{\rm rec}-\dot m$ relation, developed in this work (using \textsc{SHIVA}). From left to right, the parameters used are: $M_{\rm WD}=[1.38, 1.35, 1.25]$ $\rm M_{\odot}$, $L_{\rm WD}= 1.0 $ $L_{\odot}$, $Z=1.0 $ $Z_{\odot}$.}
    \label{fig:WDtrec}
\end{figure*}

In Table~\ref{tab:WD} we present the compiled $t_{\rm rec}$ vs $\dot m$ for RNe, which we also show in Figure~\ref{fig:WDtrec}. The $\dot M$ range from $4 \times 10^{-9}\,\rm M_{\odot}\,yr^{-1}$ to $10^{-5}\,\rm M_{\odot}\,yr^{-1}$. In the cases of KT Eri and IM Nor (as explained above), their $t_{\rm rec}$ is given as an upper limit in the first case and as a range in the second.

\begin{table*}
	\caption{List of the Galactic and extragalactic identified RNe.}
    \begin{minipage}{\textwidth}
    \centering
	\begin{tabular}{lcccccc} 
		\hline
  \footnotetext{\textbf{References}:(20)\cite{Schaefer_2010};(21)\cite{Selvelli2008};(22)\cite{patterson2017};(23)\cite{Patterson2022};(24)\cite{lederle2003};(25)\cite{Sion2019};(26)\cite{rodriguez2023};(27)\cite{schaefer2022V2487};(28)\cite{Schaefer1995};(29)\cite{thoroughgood2001};(30)\cite{stanishev2004};(31)\cite{Zamanov2023};(32)\cite{brandi2009};(33)\cite{mikolajewska2017};(34)\cite{nelson2011};(35)\cite{Schaefer2009};(36)\cite{mikolajewska2021};(37)\cite{Schaefer2022};(38)\cite{darnley2016};(39)\cite{darnley2017};(40)\cite{shore1991};(41)\cite{kuin2020};(42)\cite{bode2016}.}
		Nova & $P_{\rm orb}(\rm h)$ & Mass ($\rm{M_\odot}$) & $\dot{m}(\rm g\,cm^{-2}\,s^{-1})$ & $\dot{M}$($\rm{M_\odot}\, yr^{-1}$) & $t_{\rm rec}(\rm {yr})$ & References\\
		\hline
		T~Pyx & 1.8 & 1.25-1.40 & 0.4-1.5 & $(0.84-3.66)\times 10^{-8}$ & 19 & \citepalias{Schaefer_2010, Selvelli2008, patterson2017}\\
		IM~Nor & 2.5 & 1.0 & 1.5 & $6.6\times 10^{-8}$ & $\leq$ 82 & \citepalias{Schaefer_2010,Patterson2022}\\
		  CI~Aql & 14.8 & 1.0 $\pm$ 0.14 & 0.6 & $4\times 10^{-8}$ & 24 & \citepalias{Schaefer_2010,lederle2003,Sion2019}\\
        V2487~Oph & 18.1 & 1.35 & 3.3-24 & $(8-57)\times 10^{-8}$ & 18 & \citepalias{Schaefer_2010,rodriguez2023,schaefer2022V2487}\\
        U~Sco & 29.5 & 1.55 $\pm$ 0.24 & 0.5-0.9 & $(4-7)\times 10^{-9}$ & 10 & \citepalias{Schaefer_2010,Schaefer1995,thoroughgood2001}\\
        T~CrB & 5461 & 1.37 $\pm$ 0.13 & 0.1-1.7 & $(0.27-4.05)\times 10^{-8}$ & 80 & \citepalias{Schaefer_2010,stanishev2004,Zamanov2023}\\
        RS~Oph & 10896 & 1.2-1.4 & 0.08-1.5 & $(0.2-3.6)\times10^{-8}$ & 15 & \citepalias{Schaefer_2010,brandi2009,mikolajewska2017,nelson2011}\\
        V3890~Sgr & 12473 & 1.35 $\pm$ 0.13 & 0.4-4.2 & $(1-10)\times 10^{-8}$ & 22 & \citepalias{Schaefer_2010,Schaefer2009,mikolajewska2021}\\
        KT~Eri & 62.8 & 1.25 $\pm$ 0.03 & 3.5-34 & $(0.83-8.2)\times 10^{-7}$ & 5-50 & \citepalias{Schaefer2022}\\
        M31N~2008-12a  & 120-552 \footnote{constraints reported in \cite{darnley2017}} & 1.38 & 4-400 & $ 10^{-7}-10^{-5}$ & 1 & \citepalias{darnley2016,darnley2017}\\
        Nova~LMC~1968  & 31.2 & 1.30 & 0.04-0.4 & $10^{-9}-10^{-8}$ & 6.2 & \citepalias{shore1991,kuin2020}\\
        Nova~LMC~1971b  & 28.8 & 1.10-1.30 & 4.6-35 & $(1.1-8.3)\times 10^{-7}$ & 38 & \citepalias{bode2016} \\
		\hline
	\end{tabular}
    \end{minipage}
    \label{tab:WD}
\end{table*}

We find that the RN M31N~2008-12a \citep[with $t_{\rm rec}=1$~yr;][]{darnley2016,Henze2018} bridges the gap between the much less frequent WD eruptions ($t_{\rm rec} \sim$~decades-century) and the XRBs from NS-LMXBs ($t_{\rm rec} \sim$~minutes-months). This subcategory of RNe (with $t_{\rm rec}<10$ yr) are referred to as 'rapid recurrent novae' (RRNe) \citep{DARNLEY2020}. Only extragalactic RRNe have been observed so far (in M31), RN M31N~2008-12a being the only one with $\dot m$ estimates and thus included in Table~\ref{tab:WD} (but see also Section~\ref{subsec:extragalactic RNe} and \citealt{shafter2015}).

From our WD sample, we find no significant correlation between $t_{\rm rec}$ and $\dot m$. The Pearson correlation coefficient for our sample of 12 RNe is $c_{\rm pearson}=-0.2924$ (with a P-value of 0.3564), suggesting weak or no correlation.
This observed lack of correlation may be due to a combination of: 

\begin{enumerate}
    \item a small WD sample spanning a relatively narrow range of $\dot{m}$. Our NS sample spans 4 orders of magnitude, whereas the WD sample covers 3 orders of magnitude in $\dot{m}$. However, the $t_{\rm rec}$-$\dot m$ anti-correlation is seen in NSs over much smaller ranges ($\lesssim$1 order of magnitude, see Fig.~\ref{fig:NStrec}).

    \item Uncertainties in our estimates of $\dot{m}$. We used different methods to estimate the mass accretion rate in our WD sample, and this could increase the scatter in the $t_{\rm rec}$-$\dot m$ relation.

    \item Variance in the intrinsic parameters across our WD sample (e.g., mass and luminosity). The WD mass and base luminosity have a strong effect on recurrence time (Sec.~\ref{sec:SHIVA}). Thus, the lack of correlation seen in Figure~\ref{fig:WDtrec} could reflect different WD properties that lead to different $t_{\rm rec}$ for the same $\dot m$.

\end{enumerate}

In the future, a larger sample of RNe with measured $t_{\rm rec}$ as well as more accurate $\dot{m}$ estimates could reveal whether WDs trace a single $t_{\rm rec}-\dot m$ relation, or they trace intrinsically different tracks in the $t_{\rm rec}$-$\dot m$ plane.

\section{Ignition models}
\label{sec:Models}

\subsection{SHIVA}
\label{sec:SHIVA}

The simulations of RNe and XRBs presented in this study were conducted using the hydrodynamic, Lagrangian, finite-difference, time-implicit code \textsc{SHIVA} (see \citealt{Jose1998,jose2016stellar}, for details). 
\textsc{SHIVA} is based on the standard set of stellar evolution differential equations (conservation of mass, energy, and momentum, 
along with energy transport) and has been a extensively used in the study of stellar explosions for over 25 years. The code employs a detailed equation of state that includes contributions from a degenerate electron gas, a multicomponent ion plasma, and radiation \citep{Blinnikov1996}. Coulomb corrections to the electron pressure are accounted for, and OPAL Rosseland mean opacities are adopted \citep{Iglesias1996}. Nuclear energy generation is calculated using a reaction network comprising 120 isotopes (from $^1$H to $^{48}$Ti) and 630 nuclear processes, for recurrent novae, and 324 isotopes ($^1$H to $^{107}$Te) and 1392 nuclear interactions, for Type I X-ray bursts. Reaction rates are drawn from the STARLIB database (\citealt{Sallaska2013}, and C. Iliadis, pers. comm.). The code incorporates as well screening factors and neutrino energy losses.  

All simulations utilizing \textsc{SHIVA} in this work assume that the plasma accreted from the companion star has a Solar composition, with abundances taken
from \cite{Lodders2009}. \textsc{SHIVA} also features a time-dependent formalism for modeling convective transport, which activates when the characteristic convective timescale exceeds the integration timestep. Partial mixing between adjacent convective shells is treated using a diffusion approximation \citep{Prialnik1979}. No additional processes such as semiconvection or thermohaline mixing are included in the simulations.

\subsection{SETTLE}
\label{sec:SETTLE}

To recreate the XRB ignition models of \cite{Cumming01b} over a wide range in $\dot{m}$, we used the latest version of \textsc{pysettle}\footnote{\url{https://github.com/adellej/pysettle}} \citep{Goodwin2019}. \textsc{Pysettle} is 
a semi-analytic ignition model for XRBs, using a multi-zone model of the accreting layer and a one-zone ignition criterion. \textsc{Pysettle} assumes that the accumulating fuel layer is heated by hot CNO H burning within the layer, and a constant flux from deeper layers (see Section~\ref{sec:heat} below). It does not include heating from energy released in previous bursts, or the possibility of ignition of unburnt fuel within the ashes of previous bursts.
All \textsc{pysettle} models assume pure He accretion ($X_0$=0), and a metallicity of $Z=1.0\,Z_{\odot}$. 
Models for XRBs including compressional heating in \textsc{pysettle} were also calculated for the $t_{\rm rec}-\dot m$, but the results did not differ from the non-compressional heating models.

\subsection{Heat flux and $t_{\rm rec}-\dot m$ relation}
\label{sec:heat}

Heat flux from the NS crust \citep[e.g., due to pycnonuclear reactions,][]{Haensel90} or from the WD reaching the base of the envelope can have an important effect on $t_{\rm rec}$.
Such base flux is parameterized  in \textsc{SHIVA} by a constant luminosity imposed at the base of the envelope ($L_{\rm base}$), while \textsc{settle} uses a heat flux proportional to $\dot{M}$: the energy released per accreted nucleon ($Q_{\rm b}$).
These two parameterizations are simply related as $Q_{\rm b}=L_{\rm base}/\dot M$ (e.g.,  $L_{\rm base}$=1~$L_{\odot}$ corresponds to $Q_{\rm b}=0.25$~MeV~nucleon$^{-1}$ at $\dot{M}$ = 1\% $\dot{M}_{\rm Edd}$).

To estimate the $t_{\rm rec}-\dot m$ relation for hydrogen-rich XRBs on NSs at low accretion rates, we calculate using \textsc{SHIVA} the time until the first burst for $L_{\rm base}=$[0.1, 1.0]~$L_{\odot}$, and for three values of $\dot{M}$=[$3\times10^{-12}$, $1\times10^{-9}$, $1\times10^{-8}$]~$\rm M_{\odot}\,yr^{-1}$.
We then infer the corresponding $t_{\rm rec}$ for the resulting grid of models (assuming that the time between bursts remains constant). The resulting hydrogen XRB $t_{\rm rec}-\dot m$ relations are shown in Figure~\ref{fig:NStrec} (blue and red solid lines).

To estimate the $t_{\rm rec}-\dot m$ relation for pure He bursts on NSs at a wide range of $\dot{m}$, we run two sequences of models with \textsc{settle} using $Q_{\rm b}=1$ and 2 MeV~nucleon$^{-1}$.
From Figure~\ref{fig:NStrec} (yellow line), it is evident that the XRB ignition for pure He and $Q_{\rm b}=2$ $\rm MeV\,nucleon^{-1}$ agrees very well with the observations at $\dot{M} \lesssim$1\%$\dot{M}_{\rm Edd}$.
The two UCXBs 4U~0614+09 and 2S~0918-549, which have the most precise $t_{\rm rec}$ constraints thanks to Fermi-GBM monitoring observations \citep{Linares20124U,Jenke2016}, are connected by these $Q_{\rm b}=$2~MeV~nucleon$^{-1}$ models. 
For accretion rates $\geq~10\%$~$L_{\rm Edd}$, the observed XRB behavior is more complex (as summarized in Section~\ref{sec:NStrec}). We do not attempt to model this regime in this work, but focus instead on low-$\dot{m}$ XRBs and the connection with WD RNe.

The WD $t_{\rm rec}-\dot m$ relations (solid lines) shown in Figure~\ref{fig:WDtrec} have been modeled in this work using \textsc{SHIVA}. By using different WD masses, and keeping the same luminosity and metallicity, we obtain qualitatively different $t_{\rm rec}-\dot m$ tracks. The observed systems vary significantly in $\dot M$.
Thus, we calculate three different sequences with  $M_{\rm WD}=[1.38, 1.35, 1.25]\,\rm M_{\odot}$, all using a WD luminosity $L_{WD}=1.0\,\rm L_{\odot}$ and a metallicity $Z=1.0\,\rm Z_{\odot}$.
The resulting $t_{\rm rec}-\dot m$ models are shown in Figure~\ref{fig:WDtrec}.
As expected, only the models with the highest $\dot M$ and $M_{\rm WD}$ are able to reproduce RNe with $t_{\rm rec}$ less than a decade.

\section{Discussion}
\label{sec:Discussion}

\begin{figure*}
	\includegraphics[width=2\columnwidth]{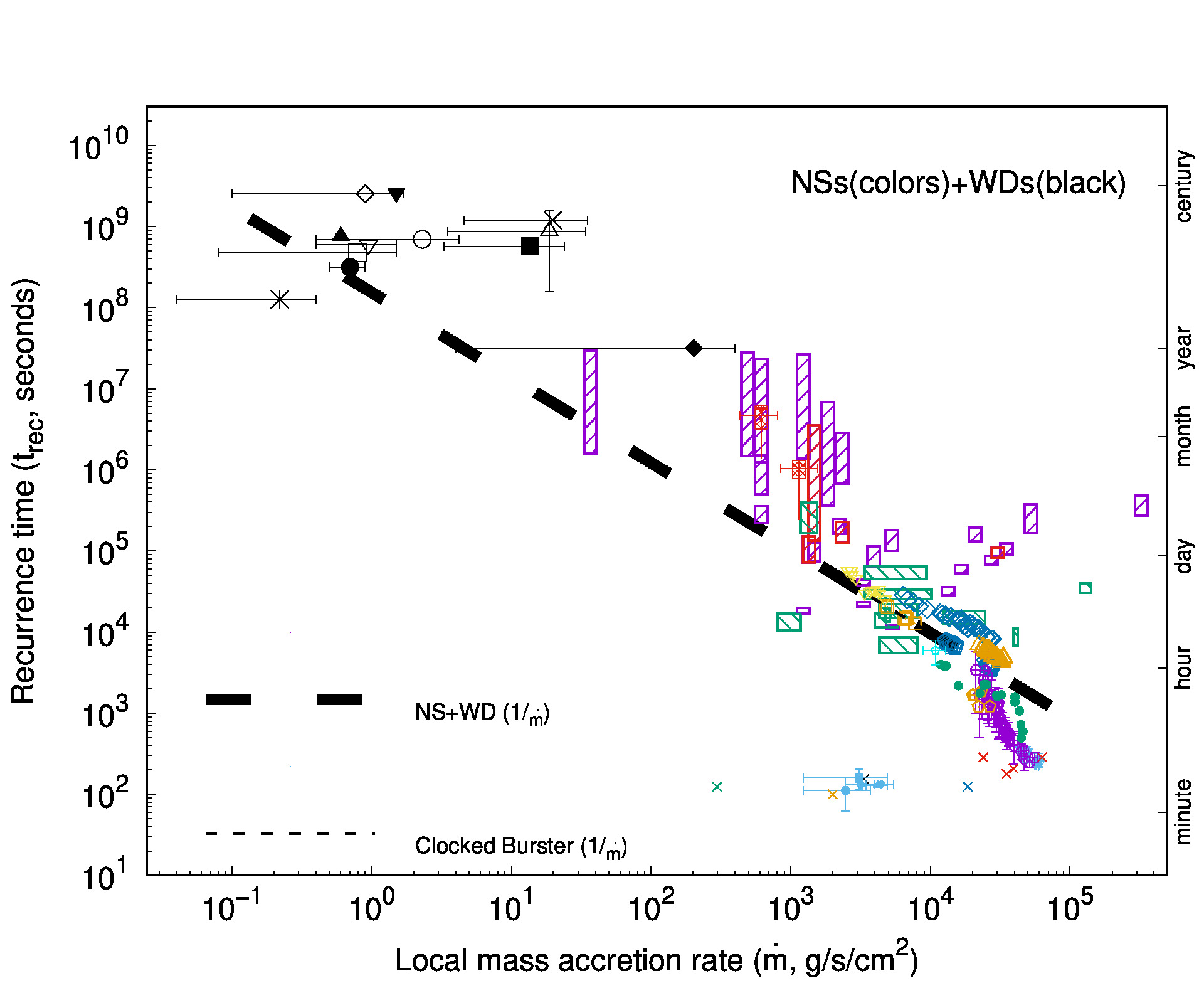}
    \caption{Recurrence times of XRBs and eruptions, and corresponding $\dot m$ on NSs and WDs. Symbols represent bursters (as in Figure~\ref{fig:NStrec}) and RNe (in black, same as Figure~\ref{fig:WDtrec}). The mHz QPOs observed in NSs are shown for comparison (as in Figure~\ref{fig:QPOs}). Lines show: $t_{\rm rec}$-$\dot{m}$ relation for the Clocked Burster (dashed black line) \protect\citep{Galloway2004}, and the connecting $1/\dot m$ relation for NS and WD (thick dashed black line).}
    \label{fig:jointtrec}
\end{figure*}

In Figure~\ref{fig:jointtrec} we bring together RNe, XRBs and mHz QPOs, in order to compare the $t_{\rm rec}$-$\dot m$ relations found in NSs and WDs.
Overall, NSs and WDs together exhibit TNRs over 5 orders of magnitude in $\dot m$ and more than 6 orders of magnitude in $t_{\rm rec}$. 
The burst/eruption energetics are very different, but both XRBs and RNe can be used as probes of thermonuclear burning in the NS and WD envelope, respectively.

We find a similar general range of mass accretion rates for our NS and WD samples: $\dot{M} \sim 10^{-10}-10^{-8}$~M$_\odot$~yr$^{-1}$ (Tables~\ref{tab:NS} and \ref{tab:WD}).
These are likely set by binary evolution processes and the long-term mass transfer rate (and limited by the Eddington limit).
However, when estimating {\it local} mass accretion rates (per unit area) we find that RNe cover $\dot m$ ranges down to 
$\dot m \sim 10^{-1}$~g~cm$^{-2}$~s$^{-1}$.
This is partly due to the much larger WD size: the accreted fuel must be spread over a larger area (by a factor $(R_{\rm WD}/R_{\rm NS})^2$).

The $t_{\rm rec}$ ranges are also different: from 1 to 80 yr in WD/RNe and between about 5~min and more than a month for NS/XRBs.
Interestingly, two systems with different accretors overlap in the $t_{\rm rec}$-$\dot m$ plane, within the large uncertainties: M31N~2008-12a and 1RXS~J171824.2-402934 (Secs.~\ref{sec:NStrec} and \ref{sec:WDtrec}, for details).
We identify these as a WD and a NS that accrete (per unit area) roughly at the same rate, and that show TNRs with about the same frequency.
We discuss their properties in more detail below in Section~\ref{sec:area}.

\subsection{NS - WD: connecting thermonuclear runaways}

Our combined NS and WD sample $t_{\rm rec}-\dot m$ relation has a Pearson correlation coefficient $c_{\rm pearson}=-0.1258$ (with $P_{\rm value}=0.3298$), indicating weak or no correlation between $t_{\rm rec}$ and $\dot m$. A power law fit of the joint sample has an index of $\alpha=-0.705$, diverting from the clocked burster relation seen in Figure~\ref{fig:jointtrec}.
We show for comparison, how a linear anti-correlation between $t_{\rm rec}$ and $\dot m$ for both NSs and WDs would look, by extrapolating in Figure~\ref{fig:jointtrec} the $1/\dot m$ relation of the clocked burster \citep{Galloway2004} to a wider range of $t_{\rm rec}$ and $\dot m$ (thick black dashed line).
%
%
As mentioned in Section~\ref{sec:NStrecRES}, the high-$\dot{m}$ XRBs from NSs show a complex phenomenology which goes beyond this simple anti-correlation. 
Moreover, the differences in fundamental system parameters (Spin Freq., $P_{\rm orb}$, fuel composition, magnetic field strength, masses) can potentially 
account for the large scatter in this power law fit. 
%

Nonetheless, this overall power law is able to connect the timescales of XRBs and RNe eruptions.
This suggests that, despite having very different sizes and magnetic field strenghts, thermonuclear runaways on WDs and NSs share similar ignition conditions.
It is worth noting that shell flashes in WDs and NSs are thought to be triggered by different nuclei: H ignition in the case of RNe and He ignition in the vast majority of XRBs.
Their recurrence times are nevertheless connected, which suggests that these are mostly set by common nuclear processes (and not by surface gravity, which is drastically different on the surface of NSs and WDs).

\subsection{Radiated energies and ignition depths for WDs and NSs}
\label{sec:area}

We now examine more closely the overlap between NS and WD $t_{\rm rec}$ and $\dot m$.
As mentioned above, M31N~2008-12a and 1RXS~J171824.2-402934 have both $t_{\rm rec}$ close to 1 yr and $\dot m$ close to $100\,\rm g\,cm^{-2}\,s^{-1}$. Thus, it is interesting to assess if the energies radiated by the TNRs in this WD and NS are consistent with the difference in radius and area between them. 

To do this, we need an estimate for the total radiated flux of M31N~2008-12a during an eruption, as well as for a burst in 1RXS~J171824.2-402934. The models of \cite{Kato2016,Kato2017} point to a total photospheric luminosity $L_{\rm phot}\sim 10^{4.7}L_{\odot}$ for M31N, leading to a total radiated eruption energy of $E_{\rm b_{\rm WD}}=5\times10^{44}$~erg (for an eruption time scale of roughly 29 d). However, this is a crude estimate, and thus can only be considered as an upper limit. The high mass of the WD implies a low critical ignition mass, meaning that a smaller amount of material is ignited than in a typical RN \citep{Henze2018}. This should result in a lower total radiated flux than the value reported here. For 1RXS~J171824.2-402934, the total radiated burst energy is estimated at $E_{\rm b_{\rm NS}}=3.0_{-1.5}^{+2.4}\,\times 10^{40}\,$~erg, for a distance of $13\pm2$ kpc \citep{Zand2005}.

With these estimates for $E_b$, we can calculate the corresponding ignition column depth $y=E_b(1+z)/4\pi R^2Q_{\rm nucl}$, where $z=\left(1/\sqrt{1-\frac{2GM}{Rc^2}}\right)-1$ is the gravitational redshift, R the NS or WD radius and $Q_{\rm nucl}=1.6+4X$ is the energy release, assuming a H fraction X at ignition \citep{Galloway2008}. For a $1.4\,\rm M_{\odot}$, 10 km radius NS, $z=0.31$. For a $1.38\,\rm M_{\odot}$, 2000 km radius WD, the redshift is $z=1\times10^{-3}$. In the case of M31N~2008-12a, we assume solar composition ($X=0.7$), and for 1RXS~J171824.2-402934, we calculate the ignition depth for both solar abundances and pure He ($X=0$). Thus, for M31N~2008-12a we derive an ignition depth $y_{\rm WD}=3.0\times10^8\,\rm g\,cm^{-2}$, whereas for 1RXS~J171824.2-402934 we get $y_{\rm NS}=7.1\times10^8\,\rm g\,cm^{-2}$ for solar abundances, and $y_{\rm NS}=1.9\times10^9\,\rm g\,cm^{-2}$ for pure He.

With the assumed radii, estimated $E_b$ and inferred y, the ratio between WD and NS areas is $(R_{\rm wd}/R_{\rm NS})^2=4\times10^4$, while the energy ratio is $E_{\rm b_{\rm WD}}/E_{\rm b_{\rm NS}}=2\times10^4$.
The inferred ignition depths, on the other hand, are similar ($y_{\rm NS}/y_{\rm WD}=(2.4-6.5)$, depending on the accreted composition).
We conclude that, apart from having similar $t_{\rm rec}$ and $\dot m$, these two sources show a similar ignition depth, and the difference in eruption/burst energy is consistent with the difference in radius and area between NSs and WDs.
In other words, the thickness of the fuel layer that ignites is similar but the WD eruption is $\sim$10$^{4}$ times more energetic since the WD radius is $\sim$10$^{2}$ times larger.


\section*{Acknowledgements}

We are grateful to J. J. M.in’t Zand and D.Galloway, for reading the manuscript and providing valuable comments that have improved this paper. We would like to thank J.Basu for his valuable insights on the M31N 2008-12a eruption energy, and M.Darnley for his critical comments. This work has been partially supported by the Spanish MINECO grant PID2023-148661NB-100, the E.U. FEDER funds, and the AGAUR/Generalitat de Catalunya grant SGR-386/2021.
ML acknowledges funding from the European Research Council (ERC) under the European Union’s Horizon 2020 research and innovation programme (grant agreement No. 101002352).

\section*{Data Availability}

The data used in this work can be provided upon reasonable request to the authors.



\bibliographystyle{mnras}
\bibliography{paper} 







\bsp	
\label{lastpage}
\end{document}